\documentclass{aa}  

\usepackage{graphicx}
\usepackage{multirow}
\usepackage{booktabs}
\usepackage{txfonts}
\usepackage{xcolor}
\usepackage[pdfpagelabels=false]{hyperref}	
\hypersetup{colorlinks=true,linkcolor=blue,citecolor=blue,filecolor=blue,urlcolor=blue,}

\defcitealias{genc_2025b}{G25}

\begin{document}

   \title{KiDS-DR5: Exploring dust transport in dark matter halos of galaxies}

   \titlerunning{Exploring dust transport in KiDS-DR5}

   \author{Eray Genc
          \inst{1}\fnmsep\thanks{\email{egenc@astro.ruhr-uni-bochum.de}}
          \and 
          Angus H. Wright\inst{1} 
          \and 
          Hendrik Hildebrandt\inst{1}
          \and 
          Dian P. Triani\inst{2,3}
          \and
          Dominik J. Bomans\inst{1,4}
          \and
          Benjamin Stölzner\inst{1}
          \and
          Marika Asgari\inst{5}
          \and
          Henk Hoekstra\inst{6}
          \and
          Benjamin Joachimi\inst{7}
          \and
          Shahab Joudaki\inst{8}
          \and
          Konrad Kuijken\inst{6}
          \and
          Lauro Moscardini\inst{9,10,11}
          \and
          Mario Radovich\inst{12}
          \and
          Robert Reischke\inst{1,13}
          \and 
          Ziang Yan\inst{14}
        }

    \institute{Ruhr University Bochum, Faculty of Physics and Astronomy, Astronomical Institute (AIRUB), German Centre for Cosmological Lensing, 44780 Bochum, Germany \and 
Institute for Theory and Computation, Harvard-Smithsonian Center for Astrophysics Cambridge, MA 02138, USA \and
Center for Astrophysics \text{\textbar} Harvard \& Smithsonian, 60 Garden Street, Cambridge, MA 02138, USA \and
Ruhr Astroparticle and Plasma Physics Center (RAPP Center), 44780 Bochum, Germany\and
School of Mathematics, Statistics and Physics, Newcastle University, Herschel Building, NE1 7RU, Newcastle-upon-Tyne, UK \and
Leiden Observatory, Leiden University, PO Box 9513, 2300 RA Leiden, the Netherlands \and
Department of Physics and Astronomy, University College London, Gower Street, London WC1E 6BT, UK \and
Centro de Investigaciones Energéticas, Medioambientales y Tecnológicas (CIEMAT), Av. Complutense 40, E-28040 Madrid, Spain
\and
Dipartimento di Fisica e Astronomia "Augusto Righi" - Alma Mater Studiorum Università di Bologna, via Piero Gobetti 93/2, I-40129 Bologna, Italy \and Istituto Nazionale di Astrofisica (INAF) - Osservatorio di Astrofisica e Scienza dello Spazio (OAS), via Piero Gobetti 93/3, I-40129 Bologna, Italy \and Istituto Nazionale di Fisica Nucleare (INFN) - Sezione di Bologna, viale Berti Pichat 6/2, I-40127 Bologna, Italy \and
INAF - Osservatorio Astronomico di Padova, via dell'Osservatorio 5, 35122 Padova, Italy \and
Argelander-Institut für Astronomie, Universität Bonn, Auf dem Hügel 71, D-53121 Bonn, Germany
\and
Kobayashi-Maskawa Institute for the Origin of Particles and the Universe (KMI), Nagoya University, Nagoya, 464-8602, Japan
}

   \date{Received XXX; accepted XXX}

 
  \abstract
   {}
   {We analyse a well-defined sample of galaxies from the final data release of the Kilo-Degree Survey (KiDS-DR5), selected via magnitude, photometric redshift, and stellar mass cuts, to measure their mean halo mass and circumgalactic dust content as a function of stellar mass. Our goal is to provide empirical constraints on dust transport mechanisms, such as stellar and AGN-driven outflows, operating in galaxy halos.}
   {We developed and validated a new measurement and modelling framework using the MICE2 simulation, implementing a realistic circumgalactic dust extinction scenario. Using this framework, we jointly modelled multiband ($ugriZYJHK_{\rm s}$) magnitude shift and galaxy–galaxy lensing signals to infer the mean halo mass and dust mass of lens galaxies at $z_{\rm B} < 0.4$, analysed in five stellar-mass bins in KiDS-DR5. The modelling jointly describes the halo mass distribution and dust extinction, including dust associated with neighbouring halos. We demonstrate that the inclusion of a satellite fraction and dust extinction from neighbouring halos is required by the data. }
   {We find a linear relation between stellar mass and dust mass, consistent with previous observational studies, but flatter than predictions from the semi-analytic model \textsc{Dusty-SAGE}. This difference may indicate that current implementations of dust-related feedback in galaxy formation models require further refinement.}
   {}


\maketitle
%

\section{Introduction}
The matter distribution in the Universe has long been a central topic of interest in astrophysics. Over the past decades, cosmologists have focused on mapping the large-scale structure of the Universe, achieving high-precision measurements through gravitational lensing techniques such as galaxy–galaxy lensing (GGL) and cosmic shear, as well as galaxy clustering \citep[see e.g. ][for reviews]{refregier_2003, kilbinger_2015}.

However, these measurements are typically limited to megaparsec scales, as quantifying the small-scale matter distribution remains challenging. On these scales, baryonic feedback processes, such as stellar winds, supernovae, and active galactic nucleus (AGN) activity, dominate, yet their impact is still poorly understood \citep{chisari_2019}.

Tracking the effects of dust grains provides a promising way to place stronger constraints on feedback models. They are produced in the interstellar medium (ISM) of galaxies, and can subsequently be transported into the circumgalactic medium (CGM) through feedback-driven outflows. Once in the CGM, some dust grains can survive for extended periods, while others are destroyed through sputtering and grain–grain collisions in the hot halo environment \citep{triani_2020}. The presence of dust in galactic halos has been observationally confirmed out to megaparsec scales from galaxy centres \citep{menard_2010, genc_2025b}.

The production, transport, and survivability of dust grains have been explored in semi-analytic and hydrodynamical simulations \citep[e.g.][]{triani_2020, richie_2024}. However, observational constraints needed to test the theoretical models in such simulations are limited. Establishing a connection between theoretical predictions and observations through measurable quantities, such as the spatial distribution, amount, and scaling of dust with galaxy properties, is essential for constraining feedback processes and understanding the evolution of circumgalactic dust.

Measuring dust in CGM is challenging due to its low surface density and the difficulty in separating the effects of circumgalactic dust extinction from those of gravitational lensing magnification, both of which alter the observed brightness of sources. In addition, the expected extinction signal is typically at the sub-percent level, requiring precise photometry, accurate background source selection, and careful control of systematic effects, such as photometric noise and intrinsic magnitude correlations.

To overcome these challenges, we adopt a joint analysis of magnitude shift (i.e. changes in observed galaxy magnitudes due to magnification and dust extinction) with multi-band photometry and GGL, which allows us to separate the contributions of dust extinction and gravitational magnification statistically. In our previous work \citep[][hereafter \citetalias{genc_2025b}]{genc_2025b}, we followed the methodology introduced by \citet{menard_2010}, who measured the average dust extinction around foreground galaxies by correlating their positions with the magnitudes of background quasars. Instead of quasars, we used background galaxies from the Kilo-Degree Survey Data Release 4 \citep[KiDS-DR4, ][]{kuijken_2019}. This leads to a lower signal-to-noise ratio (S/N) compared to quasar-based measurements, primarily due to the broader colour and magnitude distributions of galaxies, as quantified in our previous work. We validated our methodology using the MICE2 simulation \citep{fosalba_2015a, crocce_2015, fosalba_2015b}, to quantify potential systematic effects and develop mitigation strategies. While GGL provides strong constraints on the mean halo mass of the galaxy sample, multi-band photometry is essential to disentangle the wavelength-dependent dust extinction from achromatic magnification effects.

In this work, we apply our methodology to the final data release of the KiDS \citep[KiDS-DR5,][]{wright_2024} to measure the mean dust and halo mass of galaxies at photometric redshift $z_{\rm B} < 0.4$. Building on our previous study, we adopt a simulation-based framework to investigate systematic effects and to measure the dust mass\footnote{Unless stated otherwise, `dust mass' refers to the circumgalactic dust mass throughout this paper.} and the mean halo mass of galaxies. 
In this work, we extend that framework in two key ways. First, we implement a more realistic dust-extinction prescription by allowing the dust-to-stellar mass ratio to depend on galaxy properties and by explicitly accounting for the spatial extent of dust within halos. Second, we perform the measurements in bins of stellar mass and compare the resulting dust and halo mass estimates with predictions from the \texttt{Dusty-SAGE} semi-analytic model \citep{triani_2020}.


When a cosmological model is required in this analysis, we adopt the parameters from \citet{planck_2018}. The paper is organised as follows: in Sect.~\ref{sec:theo} we present the theoretical framework underlying our methodology; Sect.~\ref{sec:data} describes the MICE2 and KiDS-DR5 data set; in Sect.~\ref{sec:mice} we discuss validation tests using the MICE2 simulation; and in Sect.~\ref{sec:kids} we present our measurements of circumgalactic dust.

\section{Theoretical Framework}
\label{sec:theo}
Our aim is to infer two (population) average properties of a given galaxy sample: halo mass and halo extinction. Both quantities are estimated by correlating the angular positions of the foreground galaxies (lenses) with the features of background galaxies (sources). The lens halos modify the source shapes and fluxes in a way that encodes both halo mass and dust content in halos of lenses. In the following, we summarise the measured observables.

\subsection{Galaxy-galaxy lensing}
The mass distribution of the lenses deflects the light rays coming from sources and therefore distort their apparent shapes. This coherent distortion can be quantified statistically through the azimuthally averaged tangential shear, which is related to the excess surface mass density via
\begin{equation}
    \langle\gamma_{\rm t}\rangle(r) = \frac{\Delta\Sigma(r)}{\Sigma_{\rm crit}},
\end{equation}
where $\Delta\Sigma (r)$ is the excess surface mass density at a projected physical separation $r$. The critical surface mass density $\Sigma_{\rm crit}$ is
\begin{equation}
     \Sigma_{\rm crit} =\frac{c^2}{4\pi G}\frac{D_{\rm s}}{D_{\rm l}D_{\rm ls}}\;,
\end{equation}
where $D_{\rm l}$, $D_{\rm s}$, and $D_{\rm ls}$ denote the angular diameter distances to the lens plane, source plane, and between lens and source, respectively. The relation between critical surface density and surface mass density $\Sigma$ is expressed with the dimensionless lensing convergence:
\begin{equation}
    \kappa = \frac{\Sigma}{\Sigma_{\rm crit}}.
\end{equation}
In the weak-lensing regime, where $\kappa \ll 1$, the magnification can be approximated as
\begin{equation}
\label{eq:magn_wl}
\mu \approx 1 + 2\kappa .
\end{equation}

To infer the halo mass, we model the lens matter distribution using a Navarro–Frenk–White (NFW) density profile \citep{nfw_97}, whose parameters are linked to halo mass and redshift through a mass–concentration relation \citep{diemer_2019}. This allows us to predict the excess surface mass density $\Delta\Sigma(r)$ for a given halo mass and compare it directly with the measured GGL signal. 

\subsection{Magnitude shift}
The second observable we measure is the magnitude shift $\delta m$. We define the magnitude shift $\delta m$ as the change in the mean observed magnitude of background sources induced by foreground lenses. This change arises either from gravitational lensing by the foreground galaxies, which magnifies the flux of background sources, or from dust grains in their halos, which attenuate the observed light.
We measure the cross-correlation of the lens galaxy positions $\delta_\mathrm{g}$ with the brightness fluctuation of sources:
\begin{equation}
    \langle \delta_{\rm g}(\phi+\theta)\,\delta m(\phi) \rangle \simeq \frac{\langle \delta_{\rm g}(\phi+\theta)\,\Delta m(\phi) \rangle}{C_\mathrm{s}},
\end{equation}
where $\Delta m$ is the brightness fluctuation of the sources and defined as
\begin{equation}
    \Delta m = m - \langle m \rangle.
\end{equation}
We evaluate this signal by stacking the brightness fluctuations of background sources in annuli around the lens positions. This allows us to measure the mean magnitude shift as a function of radial separation, which represents the cross-correlation between the lens density field and the source magnitude fluctuations.
The magnitude shift consists of two contributions:
\begin{equation}
    \delta m (\lambda, \phi) = \delta m_{\mu} (\phi) + \delta m_{\rm ext}(\lambda, \phi),
\end{equation}
where $\delta m_{\mu}$ is the achromatic magnification term (i.e. the change in the observed flux of sources caused by gravitational lensing) and $\delta m_{\rm ext}$ is the chromatic extinction term caused by dust in the CGM of lenses.

The $C_\mathrm{s}$ coefficient is needed to scale the measured signal, as magnification shifts sources across the applied brightness cuts, both at the faint and bright ends. It is important to note that the $C_\mathrm{s}$ coefficient is an approximate correction to account for sources moving across the magnitude cuts, so the measured cross-correlation only approximates the true magnitude shift.

Magnification can bring sources that are intrinsically fainter than the survey limit above the detection threshold, altering the measured mean magnitude shift of the sample. This introduces a bias: the observed mean magnitude shift differs from the intrinsic one. To correct for this effect, we introduce the $C_\mathrm{s}$ coefficient \citep{menard_2010}, which scales the measured magnification signal. The $C_\mathrm{s}$ coefficient is calculated from the differential source number counts $\mathrm{d}N/\mathrm{d}m$:
\begin{equation}
    C_\mathrm{s} = 1 - \frac{1}{N_{0,\rm tot}}\frac{\mathrm{d}N}{\mathrm{d}m}(m_{\rm faint})\,[m_{\rm faint} - \langle m_0\rangle],
\end{equation}
where $\langle m_0\rangle$ and $N_{0,\rm tot}$ are the average magnitude and the total number of galaxies of the source sample after the brightness cut $m_{\rm faint}$, respectively, and the subscript $0$ indicates intrinsic values in the absence of magnification. Because the average magnification over a large sky area is unity ($\langle \mu \rangle = 1$), the global survey-wide averages of $N_{\rm tot}$ and $\langle m \rangle$ provide unbiased estimates of these intrinsic values. For sufficiently large samples, any local deviations from these means are first-order magnification effects; substituting observed for intrinsic values in the $C_\mathrm{s}$ calculation thus only introduces negligible second-order errors. Extinction-induced shifts contribute negligibly to this bias, as verified in our MICE2 simulation tests. Therefore, we adopt the formalism from \citet{menard_2010} directly.


As magnification is an achromatic effect, band-dependent effects on the measured magnitude shift signals can be attributed to the halo extinction (i.e. extinction caused by the dust in the CGM). We therefore compute the colour excess between two photometric bands $p$ and $q$:
\begin{equation}
    E_{pq} = E(\lambda_p - \lambda_q) = \delta m_p - \delta m_q\;.
\end{equation}
By measuring the color excess between two photometric bands $p$ and $q$, we isolate the dust contribution. To convert this observed reddening into a physical dust mass, we need to first convert it to extinction in $V-$band in rest-frame:
\begin{equation}
\label{eq:weighted_A_V}
A_V = \frac{E_{pq}}{k\left(\frac{\lambda_p}{1+z_{l}} \right) - k\left(\frac{\lambda_q}{1+z_{l}} \right)},
\end{equation}
where $E_{pq}$ is the reddening measured as the difference between magnitude shift signals in bands $p$ and $q$, $\lambda_p$ and $\lambda_q$ are the corresponding effective wavelengths, $k(\lambda) = A_\lambda/A_V$ is the extinction curve normalised to the $V$ band, and $z_l$ is the median redshift of the lens galaxy. 
The extinction in magnitudes is defined by the ratio of observed intensity $I$ to the intrinsic intensity $I_0$:
\begin{equation}
A_V = -2.5 \log_{10} \left( \frac{I}{I_0} \right).
\end{equation}
The intensity is attenuated by the dust optical depth $\tau_V$ ($I = I_0 e^{-\tau_V}$). Substituting this into the definition of extinction yields the relationship between magnitudes and optical depth:
\begin{equation}
A_V = -2.5 \log_{10} (e^{-\tau_V}) = \frac{2.5}{\ln 10} \tau_V \approx 1.086 \tau_V.
\end{equation}
The optical depth is related to the dust surface mass density $\Sigma_{\rm dust}$ with the absorption optical depth per unit dust mass $K_{\rm ext}(\lambda_V)$ via:
\begin{equation}
\Sigma_{\rm dust}= \frac{\tau_V(r_p)}{K_{\rm ext}(\lambda_V)} = \frac{\ln 10}{2.5} \frac{A_V}{K_{\rm ext}(\lambda_V)}.
\end{equation}
Finally, the mean dust mass in the halos of the lens galaxies is obtained by integrating the surface density profile over the impact parameter $r_p$:
\begin{equation}
M_{\rm dust}^{\rm halo} =  2\pi\int_{r_\mathrm{l}}^{r_\mathrm{u}} \Sigma_{\rm dust}(r_p)r_p  \, \mathrm{d}r_\mathrm{p} = \frac{2\pi\,\mathrm{ln}10}{2.5\,K_{\rm ext}(\lambda_V)} \int_{r_\mathrm{l}}^{r_\mathrm{u}} A_V(r_p)\,r_p\,\mathrm{d}r_{\mathrm{p}}.
\label{eq:dust_mass}
\end{equation}
where $A_V(r_p)$ is the $V$-band extinction profile. It should be noted that throughout the paper, we assume the Small Magellanic Cloud (SMC) type dust ($K_{\rm ext}=3.217\,{\rm pc}^2{M_\odot}^{-1}$), consistent with \citet{menard_2010}. 
We adopt the same integration limits as in our previous analysis, setting the lower bound to $r_{\rm l}=20\,h^{-1}\mathrm{kpc}$ to suppress contributions from dust within the interstellar medium of the lenses. The upper bound is given by the halo virial radius,
\begin{equation}
R_{\rm vir}(z)=\left(\frac{M_{\rm halo}^{\rm DM}}{\Delta_\mathrm{c}(z)\,\rho_{\rm crit}(z)}\right)^{1/3},
\end{equation}
where $M_{\rm halo}^{\rm DM}$ denotes the halo mass, $\Delta_\mathrm{c}(z)$ is the overdensity parameter, and $\rho_{\rm crit}(z)$ is the critical density of the Universe at redshift $z$.

In summary, magnification depends directly on the projected mass distribution through the surface mass density $\Sigma(r)$ and is therefore sensitive to the mean halo mass of the lens sample. In contrast, the wavelength-dependent extinction component traces the dust content of the halos and encodes the mean dust mass. As a result, the measured magnitude-shift signal responds simultaneously to both halo mass and halo extinction, and our modelling framework uses the same halo mass profile (via $\Sigma$) to describe the magnification component (as GGL) and a dust-extinction profile (via Eq.\,\ref{eq:dust_mass}) to describe the halo extinction component.

\section{Data}
\label{sec:data}
In this section, we describe the simulation data used to validate our measurement pipeline, as well as the observational data on which we estimate the mean halo mass and dust mass in stellar mass bins.
\begin{figure*}
    \sidecaption
    \includegraphics[width=12cm]{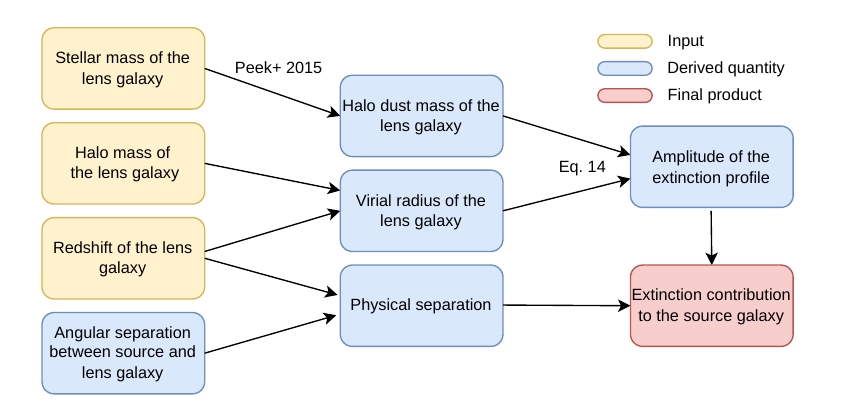}
    \caption{Simulation of circumgalactic extinction contribution to each source galaxy from one lens galaxy. Yellow colour shows the quantities already available in MICE2 catalogue, blue colour depicts derived quantities, while the red colour shows the final computed quantity.}
    \label{fig:ext_sim}
\end{figure*}

\subsection{MICE2 galaxy simulation}
\label{sec:data_mice2}
To validate our pipeline for jointly recovering the mean halo mass and dust mass of a galaxy sample from weak-lensing observables, we use the KiDS-like mock galaxy catalogue developed by \citet{jlvdb_2020}, which is based on the MICE2 cosmological simulation suite \citep{fosalba_2015a, crocce_2015, fosalba_2015b}.
MICE2 is a dark-matter–only $N$-body simulation following the growth of structure using $\sim 70$ billion particles in a box of side length $3\,h^{-1}\,\mathrm{Gpc}$ within a flat $\Lambda$CDM cosmology ($\Omega_{\rm m}=0.25$, $\Omega_\Lambda=0.75$, $h=0.7$). Halos are populated with galaxies using a combination of halo occupation distribution (HOD) and subhalo abundance matching (SHAM) prescriptions \citep{Carretero_2015}, yielding a catalogue with galaxy positions, redshifts, rest-frame magnitudes, and lensing quantities.

Lensing observables in MICE2 are derived from a convergence and shear map with a pixel resolution of $0\farcm43$. This finite resolution suppresses small-scale GGL measurements, since shape information within a pixel can partially cancel due to averaging. In contrast, magnification, being a scalar quantity, is less sensitive to sub-pixel fluctuations, allowing for measurements down to smaller angular scales (see \citetalias{genc_2025b}).

The KiDS-like mock catalogue includes intrinsic magnitudes, evolutionary corrections, magnification effects, and photometric noise. Evolutionary corrections follow \citet{fosalba_2015b}, while magnification corrections use the weak-lensing approximation (Eq.~\ref{eq:magn_wl}). These definitions allow switching magnification on or off for validation tests. However, the catalogue does not include circumgalactic dust extinction, which we therefore implement ourselves.

\subsubsection{Simulation of circumgalactic extinction with more realistic conditions}
\label{sect:dust_modelling}
In this work, we extend the MICE-based mock by incorporating a more realistic extinction model that depends on the stellar mass and size of the foreground lenses.
For each lens galaxy, we estimate its CGM dust mass using the dust–to–stellar mass relation from \citet{peek_2015}. We compute the virial radius from the halo mass and redshift of each lens galaxy and use it as a characteristic scale for the spatial extent of the dust distribution. In the following, this scale is denoted by $\beta$, such that $\beta$ effectively sets the radial extent of the extinction profile for each individual lens.

For every source galaxy, the total extinction is obtained by summing the contributions from all intervening lens galaxies along the line-of-sight. The sum runs over all lens galaxies $l$ with redshift $z_\mathrm{l}<z_\mathrm{s}$. The extinction-induced magnitude shift is therefore modelled as
\begin{equation}
\label{eq:m_ext}
\Delta m_\mathrm{ext} = \sum_{l} A_{V,l} \left( \frac{r_{\mathrm{p},l}}{\beta_l} \right)^{-0.8},
\end{equation}
where $r_\mathrm{p}$ is the projected physical separation between lens and source, and $\beta$ denotes the virial radius of the corresponding lens galaxy, setting the characteristic scale of the extinction profile. The power-law slope of $-0.8$ is motivated by the observational findings of \citet{menard_2010}, who measured a similar radial dependence of dust extinction around galaxies. This formulation implicitly assumes azimuthal symmetry of the dust distribution around lens galaxies, such that the extinction depends only on radial distance from the lens centre.

The normalisation factor $A_{V,l}$ encapsulates the overall amplitude of the extinction signal in $V$-band. The extinction contribution to other wavelengths is calculated by assuming a $1/\lambda$ scaling for the extinction law. The extinction in a specific band $i$ is given by:
\begin{equation}
A_i = A_V \left( \frac{\lambda_V}{\lambda_i} \right),
\end{equation}
where $\lambda_i$ is the effective wavelength of the corresponding filter.

In our simulation framework, we therefore assume that all galaxies follow this dust distribution, adopting a common radial slope while allowing the overall normalisation to vary. By defining the impact parameter $\beta$ through the halo mass and redshift of each individual lens galaxy, the spatial extent of the extinction profile naturally varies from system to system, providing a more realistic spatial distribution of dust for a given total dust mass. This mass- and redshift-dependent scaling was not included in our previous simulation framework.
The workflow for computing the extinction contribution from each foreground lens is illustrated in Fig.~\ref{fig:ext_sim}. 

\subsection{KiDS-DR5}
The Kilo-Degree Survey (KiDS) is a wide-field imaging survey conducted with the 2.6\,m VLT Survey Telescope \citep[VST;][]{capaccioli_2005} at the European Southern Observatory’s Paranal site in Chile. Observations are carried out with the OmegaCAM wide-field imager \citep{kuijken_2011}, which provides a $1^\circ \times 1^\circ$ field of view with a pixel scale of $0\farcs213$. The fifth and final data release, KiDS-DR5 \citep{wright_2024}, covers a total of 1347\,deg$^2$ in ten optical and near-infrared (NIR) bands ($ugri_1i_2ZYJHK_{\mathrm{s}}$), combining VST optical imaging with VIKING NIR observations \citep{edge_2013}. The survey reaches a median $5\sigma$ depth of $r \simeq 24.8$, and the effective area suitable for weak-lensing analyses after masking is approximately 967.4\,deg$^2$ \citep{wright_2024}.

Multi-band photometry is measured using the Gaussian Aperture and PSF (\texttt{GAaP}) method \citep{kuijken_gaap, kuijken_2015, kuijken_2019}, which provides accurate and PSF-homogenised colour measurements across all ten filters. KiDS-DR5 contains approximately 41 million galaxies with reliable shape measurements obtained using the \texttt{lensfit} algorithm. After quality cuts and weighting, this corresponds to a source number density of $8.81$ galaxies per arcmin$^2$, with sources spanning the photometric-redshift range $0.1 < z_B < 2.0$.

Stellar masses for KiDS-DR5 galaxies are derived using the methodology described in \citet{wright_2019}, based on spectral energy distribution (SED) fitting to multi-band optical and near-infrared photometry. Specifically, template-based SED fitting is performed using the \textsc{LePhare} code, adopting photometric redshift estimates as fixed inputs. The modelling assumes stellar population synthesis templates from \citet{bruzal_2003}, the initial mass function from \citet{chabrier_2003}, and parametrized star formation histories, together with the dust attenuation law of \citet{calzetti_1994}. Stellar masses are then obtained by scaling the best-fitting template to the observed photometry.

Photometric redshifts are estimated using the Bayesian Photometric Redshift code \texttt{BPZ} \citep{benitez_2000}, and redshift distributions for our various samples of galaxies are estimated via self-organising maps using the methods of \citet{wright_2025}. This redshift distribution estimation method relies on 27\,deg$^2$ of spectroscopic reference samples, and is demonstrated to recover the underlying true mean of faint galaxy samples\footnote{\citet{wright_2025} analyse the weak-lensing `KiDS-Legacy' sample that is an overlapping (but not perfect) sub- or super-set of the sources used here.} with percent-level accuracy and sub-percent level precision.  This improved $N(z)$ estimation method is a key feature of KiDS-DR5, and significantly enhances its suitability for precision cosmological and galaxy-evolution analyses. As such, KiDS-DR5 presents an ideal dataset for our investigation: it delivers accurate multi-band photometry needed for measuring wavelength-dependent magnitude shifts, precise galaxy shape measurements for GGL, and a robustly estimated source redshift distribution. 

\begin{figure}
    \centering
    \includegraphics[width=0.9\linewidth]{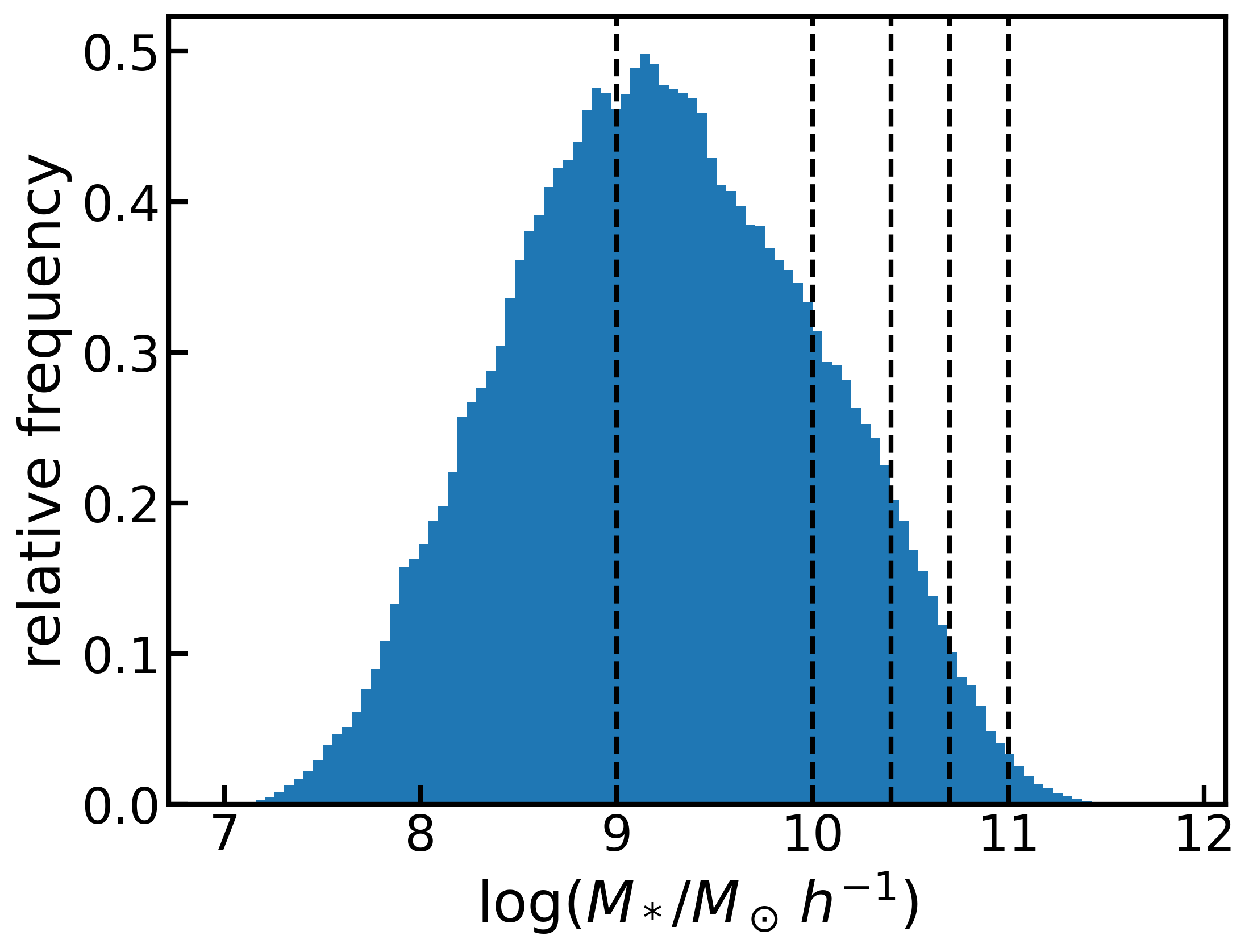}
    \caption{Stellar mass distribution of our lens sample in MICE2 analysis.
    }
    \label{fig:mice_lms}
\end{figure}

\section{Tests with the MICE2 simulation}
\label{sec:mice}
As mentioned above, our aim in this work is to measure the circumgalactic dust mass of galaxies as a function of stellar mass. For that, we extend the methodology presented in our previous study \citepalias{genc_2025b}. In Sect.~\ref{sec:mice_model}, we describe how the magnitude shift and GGL signals are measured, while in Sect~\ref{sec:mice_signal_modelling} we describe our framework to model the measured signals. Finally, we present the set-up of the parameter estimation in Sect.~\ref{sec:mice_mcmc} and discuss the results.

\subsection{Measurements}
\label{sec:mice_model}

In our previous work \citepalias{genc_2025b}, we validated our measurement pipeline designed to jointly infer the mean dust mass and mean halo mass of a galaxy sample, using multi-band magnitude shift measurements in combination with GGL signals, within the MICE2 simulation. We demonstrated that both observables can be robustly recovered in the presence of realistic photometric noise and redshift uncertainties. That analysis focused on a single galaxy population and adopted a simplified dust extinction prescription. In this work, we extend the MICE2 validation to enable measurements of dust mass as a function of stellar mass, which required a more complex modelling framework and additional free parameters. Our stellar-mass-dependent dust extinction modelling in this analysis is described in Sect. \ref{sect:dust_modelling}. 

Firstly, we select galaxies from the MICE2 catalogue covering the same sky area as KiDS-DR5, corresponding to approximately $1350\,\mathrm{deg}^2$. Source galaxies are defined by the criteria $z_{\rm B} > 0.7$ and $m_x < m_x^{\rm lim}$, where $x$ denotes the photometric band and $m_x^{\rm lim}$ the corresponding magnitude limit. Lens galaxies are selected with $z_{\rm B} < 0.42$, $m_r < 23$, and are divided into five bins of stellar mass as given in Table~\ref{tab:mice_mass_bins}. All magnitude cuts are applied to observed magnitudes that include the effects of photometric noise, magnification, and dust extinction, while photometric redshift estimates based on KiDS-DR5 data are used for the redshift selection.  We measure the magnitude shift and GGL signals for five different stellar mass bins of lenses. An overview of the lens galaxy samples is first shown in Fig.~\ref{fig:mice_lms}, which presents the stellar mass distribution of the lenses. Some stellar-mass bins - those at the lower mass end - are intentionally broader than others in order to approximately match the S/N across bins. Summary statistics of these lens populations are then provided in Table~\ref{tab:mice_mass_bins}, which lists the median stellar mass and median redshift for each stellar-mass bin.

It is important to note that the maximum redshift of KiDS-DR5 reaches $z \simeq 2.0$, whereas the MICE2 simulation extends only to $z \simeq 1.4$, therefore our source galaxy sample in MICE2 represents a subset of the full KiDS-DR5 source population. However, only a small fraction of the KiDS–DR5 sources ($\sim3\%$) lie at $z>1.4$, such that the vast majority of the source redshift distribution is common to both datasets. Although high-redshift sources are expected to have less reliable photometric redshifts, we do not expect them to introduce a systematic bias in our measurements, due to their small relative contribution. 

Using \textsc{TreeCorr} \citep{treecorr}, we measure the cross-correlation between lens galaxy positions and the magnitude shifts of background galaxies, as well as the shear of sources for GGL signal. The magnitude shift signal is measured in eight logarithmically spaced angular bins between 0.1 and 60 arcmin, while the GGL signal is measured in ten bins spanning 1 to 60 arcmin. Fewer bins are used for the magnitude shift to compensate for its lower S/N and to maximise the S/N in each bin. Angular scales below 1 arcmin are excluded from the GGL analysis, as they approach the resolution limit of the MICE2 lensing maps and result in biased shear estimates (see Sect.~\ref{sec:data_mice2}).

For each observable, we apply a random subtraction procedure to remove spurious correlations arising from residual systematic effects \citep{singh_2017}. Uncertainties are estimated using the built-in covariance matrix calculation in \textsc{TreeCorr}, adopting a jackknife resampling with 1300 spatial patches each being 1 deg$^2$ (the resulting measurements with error bars are shown in Fig.~\ref{fig:mice_dustprofiles}).

\subsection{Modelling of the measured signals}
\label{sec:mice_signal_modelling}
\begin{table}
\caption{Stellar mass bins in MICE2 lens galaxy sample with summary statistics.}
\label{tab:mice_mass_bins}
\centering
\begin{tabular}{lcccc}
\hline
Bin & Mass Range & $M_*^{\rm med}$&$z_{\rm B} ^{\rm med}$ & $N_\mathrm{tot}$ \\
\hline
Bin 1 & (9.0,10.0] & 9.42 & 0.303 & 2579691  \\
Bin 2 & (10.0,10.4]& 10.19 & 0.304 & 592870 \\
Bin 3 & (10.4,10.7]& 10.53 & 0.305 & 253969 \\
Bin 4 & (10.7,11.0]& 10.82 & 0.309 & 107000 \\
Bin 5 & $>11$& 11.11 & 0.295 & 29163 \\
\hline
\end{tabular}
\tablefoot{Stellar mass values are given in units of $\log(M_*/M_\odot\,h^{-1})$, while the superscript `med' denotes the median value of the sample.}
\end{table}

As a next step in the MICE2 tests, we infer the physical properties of the lens galaxy samples through modelling of the measured signals. While our primary parameters of interest are the mean halo mass and the dust mass of each lens sample, a realistic description of the underlying (dust) mass distribution is required to obtain robust parameter estimates. We therefore model the observed magnitude shift and GGL signals within a unified framework that accounts for fraction of satellite galaxies, (dust) mass distribution in one halo, and large-scale contributions.

A complete model of the observed magnitude shift should include contributions from both magnification and dust extinction, while GGL provides an independent probe of the underlying mass distribution of galaxies and therefore plays a crucial role in breaking the degeneracy between magnification and dust extinction when the two are modelled jointly.

\begin{figure*}
    \centering
    \includegraphics[width=0.9\linewidth]{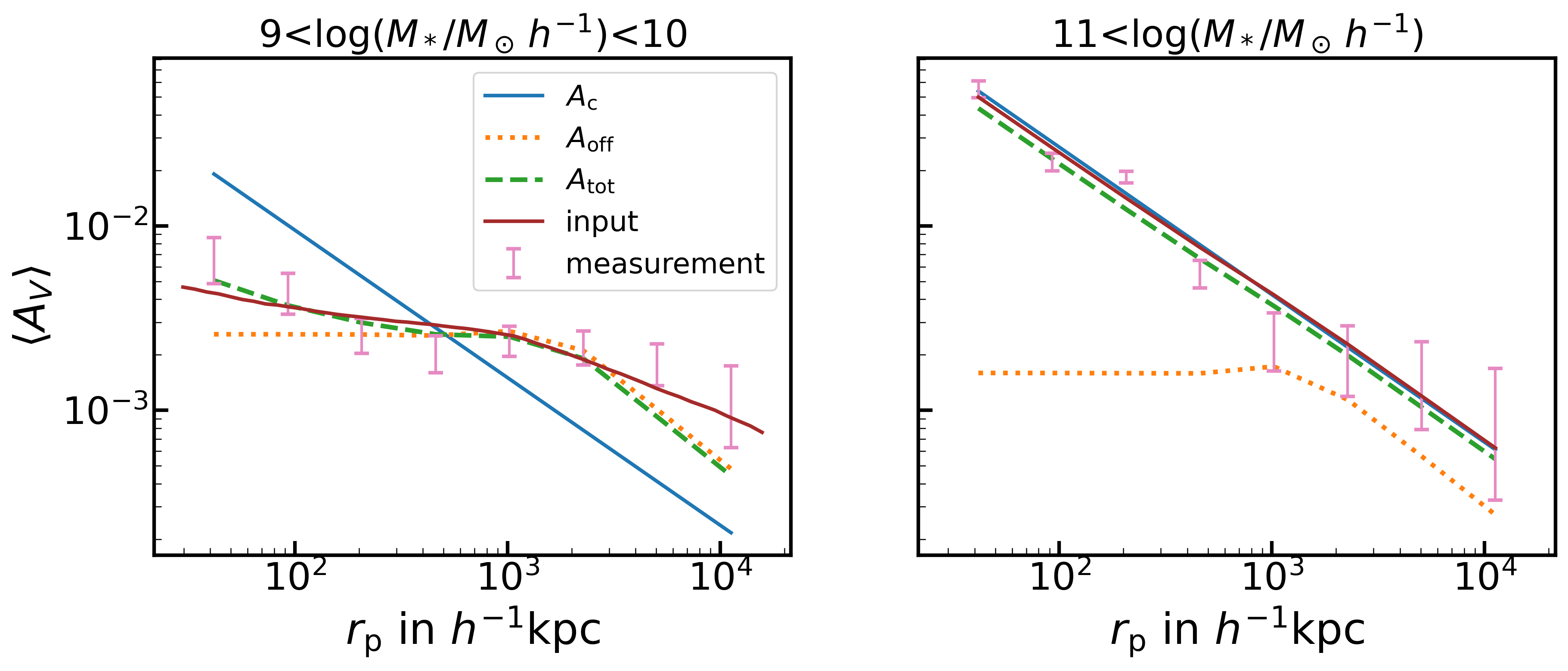}
    \caption{MICE2 dust extinction profiles for lowest and highest stellar mass bins. The left panel shows the results for the lowest stellar mass bin, while the right panel corresponds to the highest stellar mass bin. In both panels, the blue solid line represents the 1-halo extinction profile associated with central galaxies, the pink data points indicate the measured extinction signal, and the dotted lines show the off-centred host halo contribution for satellite galaxies. The green dashed lines depict the total model extinction profile, while the dark red solid line denotes the input mean extinction profile used in the MICE2 dust simulations. 
    }
    \label{fig:mice_dustprofiles}
\end{figure*}

The magnification and GGL signals are modelled within the halo model framework, assuming that galaxies reside in dark matter halos described by a Navarro–Frenk–White (NFW) density profile \citep{nfw_97}. The density profile of a single halo (the so-called 1-halo term) is given by
\begin{equation}
\rho(r) = \frac{\rho_{\rm s}}{\left( r / r_{\rm s} \right)\left(1 + r / r_{\rm s}\right)^2},
\end{equation}
where $\rho_{\rm s}$ and $r_{\rm s}$ denote the characteristic density and scale radius, respectively. 
In addition to the 1-halo contribution, correlated large-scale structure around the lens galaxies contributes to the signal through the so-called 2-halo term. We model this component using the analytic infalling profile implemented in the \textsc{Colossus} framework \citep{diemer_2018}, assuming a flat $\Lambda$CDM cosmology consistent with \citet{planck_2018} parameters. The corresponding density profile of this analytic infalling profile is expressed as
\begin{equation}
\label{eq:rho_2h}
\rho(r) = \delta_1\,\rho_{\rm m}(z)\,
\left[
\left(\frac{\delta_1}{\delta_{\rm max}}\right)^2 +
\left(\frac{r}{r_{\rm pivot}}\right)^{s/0.5}
\right]^{-1/2},
\end{equation}
where $\rho_{\rm m}(z)$ is the mean matter density at redshift $z$, $\delta_1$ sets the normalisation at the pivot radius (taken to be $R_{200}$), and $s$ controls the slope of the profile. Here, $\delta_{\max}$ sets the asymptotic inner overdensity of the infall profile, such that the density approaches $\rho(r \to 0) = \delta_{\max}\,\rho_{\rm m}(z)$.

While the above formalism was sufficient in our previous work, splitting the lens sample into stellar mass bins leads to strongly varying satellite fractions, particularly in the lowest stellar mass bins. In these low-mass bins, the satellite fraction is significantly higher than in the higher-mass bins, and the observed signal is dominated by the host halos of satellite galaxies rather than by the satellites' own halos. It is therefore necessary to explicitly model the satellite contribution.

Following \citet{sifon_2015}, we describe the total projected surface mass density as a mixture of central and satellite components:
\begin{equation}
\label{eq:sifon}
\Sigma(R) =
(1 - f_{\rm sat})\,\Sigma_{\rm 1h}(R)
+ f_{\rm sat}\,\Sigma_{\rm host}^{\rm off}(R)
+ \Sigma_{\rm 2h}(R),
\end{equation}
where $f_{\rm sat}$ denotes the satellite fraction of the lens sample. The first and third terms correspond to the standard 1-halo and 2-halo contributions, while $\Sigma_{\rm host}^{\rm off}$ accounts for the host halo contribution to satellite galaxies.
The host contribution is modelled by convolving the central halo profile with an off-centring distribution that describes the projected distance between the satellite galaxy and the centre of its host halo. Assuming a fixed satellite offset radius $R_{\rm sat}$, the off-centred profile is given by
\begin{equation}
\label{eq:sifon_avg}
\Sigma_{\rm host}^{\rm off}(R) =
\frac{1}{2\pi}
\int_0^{2\pi}
\Sigma_{\rm host}\!\left(
\sqrt{R^2 + R_{\rm sat}^2 + 2 R R_{\rm sat}\cos\theta}
\right)\,{\rm d}\theta ,
\end{equation}
which is evaluated numerically via angular averaging.

Dust extinction is modelled separately and added to the magnification signal to obtain the total magnitude shift signal. 
In analogy with the halo mass modelling described above, we account for contributions from both central and satellite galaxies by explicitly modelling the dust distribution within individual halos and the host halo contribution to satellites. The total extinction profile is written as
\begin{equation}
\label{eq:dust_modelling_tot}
A_{\rm tot}(R) =
(1 - f_{\rm sat})\,A_{\rm c}(R; M_{\rm d,1}) + 
f_{\rm sat}\,A_{\rm host}^{\rm off}(R; M_{\rm d,2}),
\end{equation}
where $A_{\rm c}(R; M_{\rm d,1})$ denotes the extinction profile associated with central galaxies, and $A_{\rm host}^{\rm off}(R; M_{\rm d,2})$ represents the host halo contribution.
In analogy with the mass modelling, the off-centred extinction profile for satellites is computed using the same angular averaging formalism as in Eq.~\eqref{eq:sifon_avg}:
\begin{equation}
\label{eq:dust_avg}
A_{\rm host}^{\rm off}(R) =\frac{1}{2\pi}\int_0^{2\pi}A_{\rm host}\left(\sqrt{R^2 + R_{\rm sat}^2 + 2 R R_{\rm sat}\cos\theta}\right){\rm d}\theta,
\end{equation}
where $A_{\rm host}$ is the radial extinction profile of the host halo, normalised by the dust mass $M_{\rm d,2}$. It is important to note that the two profiles in Eq. \eqref{eq:dust_modelling_tot} have independent dust masses, $M_{\rm d,1}$ and $M_{\rm d,2}$, where $M_{\rm d,2}$ quantifies the contribution of dust extinction from neighbouring halos on satellites.

The radial shape of the extinction profile for each halo is modelled using Eq.~\eqref{eq:m_ext}, with a fixed slope of $-0.8$, consistent with the dust distribution adopted in our MICE2-based extinction simulations (Sect.~\ref{sect:dust_modelling}). While the slope is fixed, the spatial extent of the profile $\beta$ is set by the halo virial radius, which is computed from the halo mass, $M_{\rm c}$, and redshift, thereby determining the radial distribution of dust. The amplitude, $A_0$, is not treated as a free parameter itself but is calculated from the total dust mass of the halo using Eq.~\eqref{eq:dust_mass}.

In summary, our model consists of the following free parameters: the halo mass, $M_{\rm c}$; the satellite fraction, $f_{\rm sat}$; and the parameters of the two-halo term, $\delta_1$ and $s$, which enter Eq.~\eqref{eq:rho_2h} as described above; together with the dust masses, $M_{\rm d,1}$ and $M_{\rm d,2}$, which set the amplitude of the extinction profiles for central and neighbouring halos, respectively. 

Fig.~\ref{fig:mice_dustprofiles} illustrates the impact of satellite host halo contributions on the recovered extinction profiles in the MICE2 validation. We show the mean extinction profile (rest-frame extinction converted to the $V-$band) derived from the nine-band $ugriZYJHK_{\rm s}$ colour contributions in the MICE2 simulations for the lowest and highest stellar mass bins only, in order to highlight the contrast between these regimes and enable a direct visual comparison.

The profiles are computed as a weighted average of the reddening signals across the optical $ugri$ bands, as well as the five near-infrared bands $ZYJHK_s$. For each band combination (excluding combinations between different NIR bands, which were found to be negligible), the extinction is calculated from the reddening signal using Eq.\eqref{eq:weighted_A_V}.
The final extinction profile for each stellar-mass bin is then obtained as a covariance-weighted average over all band combinations,
\begin{equation}
\label{eq:Av_mean}
\langle A_V \rangle = \frac{\sum_{i,j} A_V^{(i)} \, (C^{-1})_{ij}}{\sum_{i,j} (C^{-1})_{ij}},
\end{equation}
where $A_V^{(i)}$ denotes the extinction estimate from the $i$-th band combination and $C$ is the covariance matrix between all band combinations, thereby properly accounting for correlations between measurements.

For the lowest stellar mass bin, the inclusion of the off-centred host contribution is essential to reproduce the measured extinction signal on large scales, particularly beyond $\sim300\,h^{-1}\,\mathrm{kpc}$. This behaviour reflects the high satellite fraction in this bin, for which the observed extinction profile is dominated by dust associated with neighbouring host halos rather than by the satellites’ own halos. As a result, the mean extinction profile of this sample is significantly flatter than expected from a single 1-halo dust model, and cannot be adequately described by $A_{\rm c}$ alone. In contrast, for the highest stellar mass bin, the extinction signal is largely driven by the galaxies’ own halos, and the contribution from off-centred host halos is subdominant. Consequently, the total extinction profile closely follows the central 1-halo term, and the inclusion of the satellite host component has only a minor effect on the recovered profile. Measurements for the intermediate stellar mass bins and their best-fitting models are presented in Fig.~\ref{fig:mice_vs_kids_Av}.


\subsection{Parameter estimation}
\label{sec:mice_mcmc}

To infer the physical parameters of the lens galaxy samples from the measured magnitude shift and GGL signals, we employ a Bayesian parameter estimation framework based on Markov Chain Monte Carlo (MCMC) sampling. This approach allows us to efficiently explore the multi-dimensional parameter space of our model and to quantify parameter degeneracies and uncertainties in a self-consistent manner.

We define a data vector $\boldsymbol{\xi}$ that consists of the concatenated magnitude shift measurements in the four photometric bands and the GGL measurements. For a given set of model parameters, the corresponding model prediction is denoted by $\boldsymbol{\xi}_{\rm m}$. Assuming Gaussian-distributed measurement uncertainties, the likelihood is given by
\begin{equation}
\label{eq:likelihood}
    \ln\mathcal{L} \propto -\frac{1}{2} \left[ \boldsymbol{\xi} - \boldsymbol{\xi}_m \right]^\intercal \mathbf{C}^{-1} \left[ \boldsymbol{\xi} - \boldsymbol{\xi}_m\right]
\end{equation}
where $\mathbf{C}$ is the full covariance matrix estimated from jackknife resampling.
Sampling of the posterior distribution is performed using the affine-invariant ensemble sampler \textsc{emcee} \citep{emcee}. We jointly fit the four-band ($ugri$) magnitude shift measurements and the GGL signal using a model with seven free parameters described above.

Uniform priors are adopted over physically motivated ranges for all parameters; the adopted prior intervals are summarised in Table~\ref{tab:priors}. For each stellar mass bin, we initialise 50 walkers and run the sampler for a total of 2000 steps, discarding the first 300 steps as burn-in and using the remaining samples to estimate the posterior distributions. Convergence was assessed through visual inspection of the chains and through calculation of the Gelman-Rubin scale-reduction factor. We confirmed that $\hat{R}<1.05$ for each signal, ensuring that the multiple MCMC chains reached a common stationary distribution.

In Table~\ref{tab:results_mice2}, we present the best-fit parameters for the mean halo mass ($M_{\rm c}$) and dust mass ($M_{\rm d,1}$) within the $1\sigma$ confidence interval for each stellar-mass bin, alongside the true values in the simulation. The table also reports the $\chi^2$ and probability-to-exceed (PTE) values for each measurement. These results are obtained from the MCMC analysis of the joint measurement of GGL and four-band magnitude shift signals. Best-fit values correspond to the maximum-a-posteriori (MAP) estimates, while the quoted uncertainties represent the 68\% projected joint-distribution highest posterior density (PJ-HPD) intervals. Overall, the MCMC is able to recover the input parameters accurately across the full stellar-mass range.

\begin{figure*}
    \centering
    \includegraphics[width=0.9\linewidth]{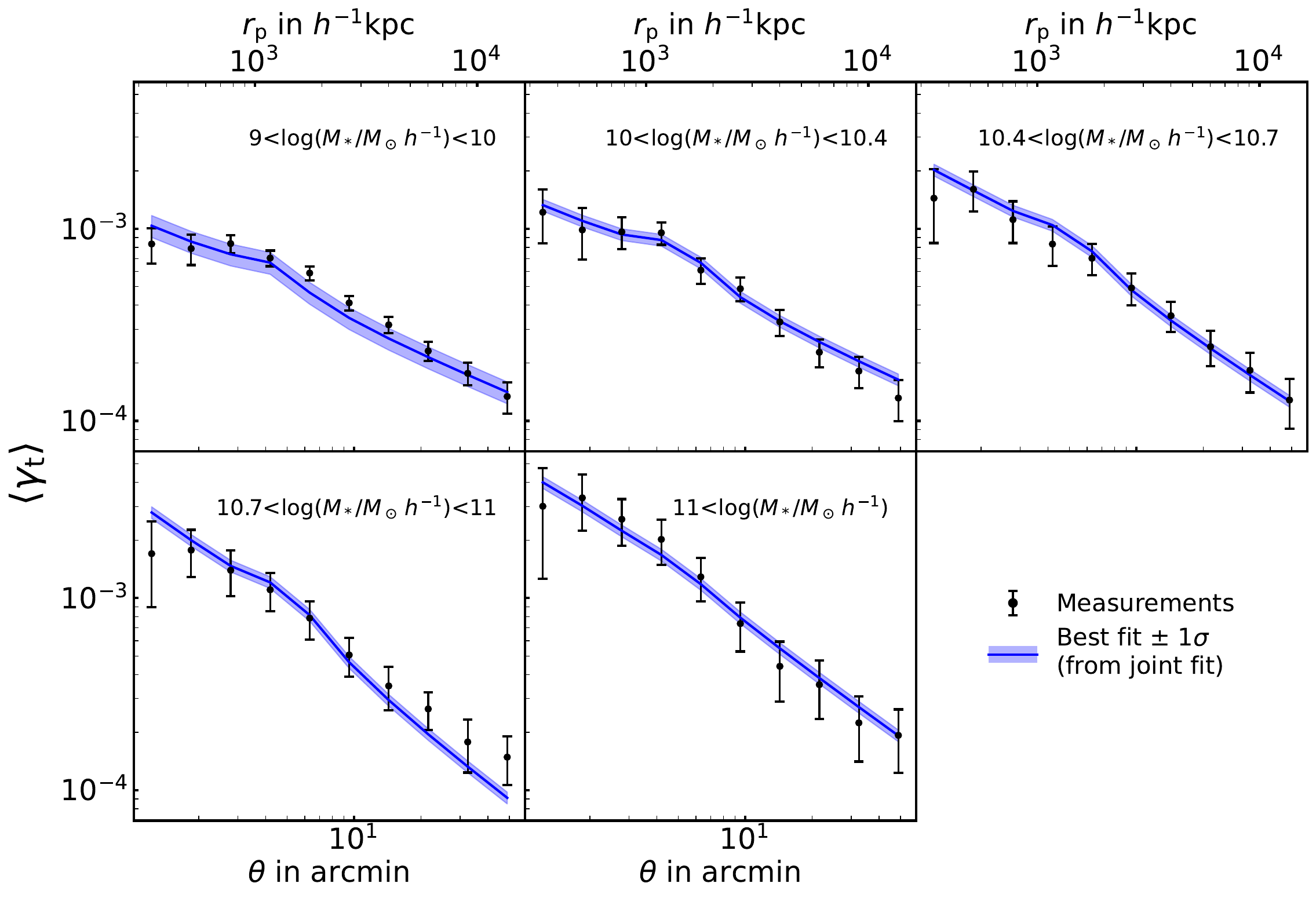}
    \caption{GGL signals in stellar mass bins measured in the MICE2 data, together with the best-fit model and its $1\sigma$ posterior credible interval from a joint fit with the magnitude shift measurements. The conversion between angular and physical separations assumes the median redshift of the sample, $z\sim0.3$, which is approximately the same for all stellar-mass bins.}
    \label{fig:mice_ggl}
\end{figure*}

Figure~\ref{fig:mice_ggl} presents the GGL measurements in five stellar mass bins, each shown in a separate panel together with the best-fitting model and its $1\sigma$ confidence interval. The model curves are obtained from the joint analysis of GGL and magnitude shift signals. The error bars on the data points correspond to the square roots of the diagonal elements of the full covariance matrix, i.e. the marginal uncertainties. Overall, the best-fitting models are in good agreement with the measured signals across all mass bins. In the lower stellar mass bins, we observe mild excess features (“bumps”) on intermediate scales, which reflect the enhanced contribution from host halos of satellite galaxies. The GGL signal is particularly important for constraining the satellite fraction $f_{\rm sat}$ and the characteristic satellite offset scale $R_{\rm sat}$, both of which are essential inputs for the dust extinction modelling.

\begin{figure*}
    \centering
    \includegraphics[width=0.9\linewidth]{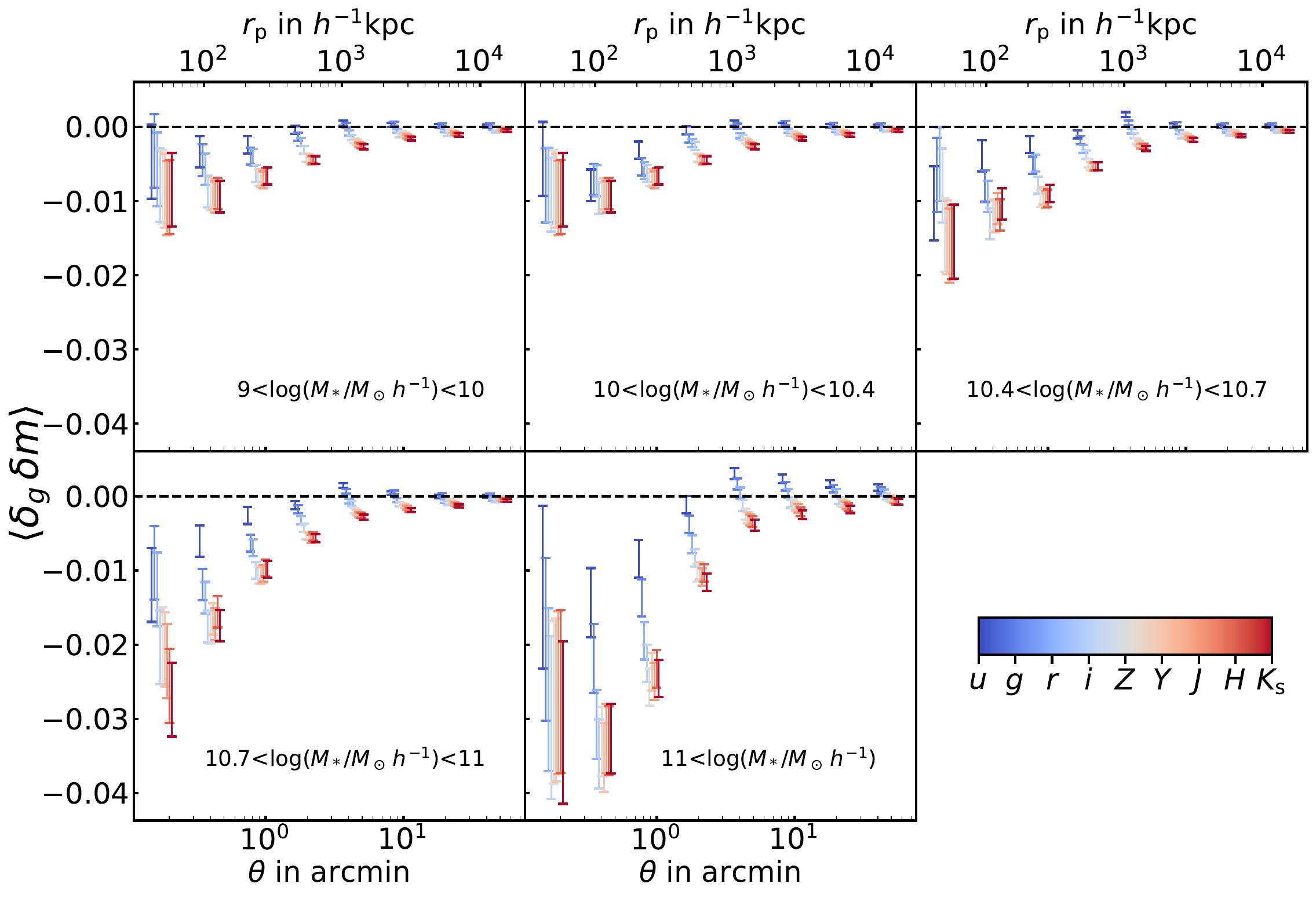}
    \caption{Magnitude-shift signals measured in MICE2 galaxies across different $ugriZYJHK_{\rm s}$ bands. Physical separations are computed assuming the median redshift of the lens samples, $z\sim0.3$.}
    \label{fig:mice_magn}
\end{figure*}

Figure~\ref{fig:mice_magn} shows the corresponding magnitude shift measurements, again split into five stellar mass bins. In each panel, the nine-band $ugriZYJHK_{\rm s}$ magnitude shift measurements are shown with colours ranging from blue to red to indicate increasing wavelength. For clarity, the best-fitting models are not shown in this figure. Instead, Fig.~\ref{fig:mice_vs_kids_Av} presents the rest-frame mean extinction profiles derived from the magnitude shift measurements together with the MCMC best-fitting models, allowing the goodness of fit to be assessed directly.

As expected, the amplitude of both GGL and magnitude shift signals generally increases with stellar mass. It is important to note that negative magnitude shifts correspond to brighter observed sources. We notice that $u$-band signal becomes even positive on large scales for higher stellar mass bins. This feature arises from the high dust mass in these galaxies, combined with the fact that the $u$-band is the most sensitive to extinction due to its short wavelength. In all stellar mass bins, the best-fitting models provide a good description of the observed signals, as evidenced by PTE values lying within the acceptable range of 0.05–0.95. These tests indicate that our joint modeling framework is able to robustly recover the true halo and dust properties, validating its reliability for application to observational data, the analysis of which is presented in the following section.

\begin{table}
    \centering
    \caption{Free parameters and the physically motivated priors used in the MCMC analysis.}
    \begin{tabular}{c c l}
        parameter & prior & description \\ 
        \hline \vspace{1.6mm}
        $\log M_{\rm c}$ & [10, 15] 
        & Halo mass \\ \vspace{1.7mm}
        $f_{\rm sat}$ & [0.1, 1.0] 
        & Satellite fraction \\ 
        $\delta_1$ & [1, 100]  
        & Normalisation of the 2-halo term \\ 
        $s$ & [0.1, 4]
        & Slope of the 2-halo term \\
        $R_{\rm sat}$ & [0, 4]
        & Off-centring radius of satellites\\
        $\log M_{\rm d,1}$ & [6, 10] 
        & Central halo dust mass \\
        $\log M_{\rm d,2}$ & [6, 10] 
        & Satellite halo dust mass \\
        \hline
    \end{tabular}
    \label{tab:priors}
    \tablefoot{Halo and dust mass parameters are expressed in $\log(M/M_\odot\,h^{-1})$, and all radial quantities are given in $\mathrm{kpc}\,h^{-1}$.}
\end{table}

\begin{table*}
\centering
\caption{Best-fit parameters from the joint magnification, dust extinction, and GGL analysis in five stellar-mass bins in MICE2.}
\label{tab:results_mice2}
\begin{tabular}{lccccc}
\toprule
Parameter & Bin 1 & Bin 2 & Bin 3 & Bin 4 & Bin 5 \\
\midrule
$\log(M_{\mathrm{c}})$ & $11.72^{+0.17}_{-0.18}$ & $11.80^{+0.11}_{-0.10}$ & $12.16^{+0.12}_{-0.09}$ & $12.29^{+0.10}_{-0.09}$ & $12.58^{+0.11}_{-0.10}$ \\ [0.4em]
Simulation & 11.56 & 11.72 & 12.11 & 12.35 & 12.60 \\
\midrule
$\log(M_{\mathrm{d1}})$ & $7.49^{+0.09}_{-0.11}$ & $7.71^{+0.11}_{-0.11}$ & $7.70^{+0.13}_{-0.12}$ & $7.84^{+0.10}_{-0.11}$ & $8.08^{+0.08}_{-0.09}$ \\ [0.4em]
Simulation & 7.36 & 7.68 & 7.81 & 7.93 & 8.05 \\
\midrule
$\chi^{2}\;(\nu=35)$ & 45.85 & 42.66 & 43.79 & 41.83 & 42.43 \\
PTE & 0.10 & 0.17 & 0.15 & 0.20 & 0.18 \\
\bottomrule
\end{tabular}
    \tablefoot{$M_{\mathrm{c}}$ denotes the mean host halo mass and $M_{\mathrm{d1}}$ the mean dust mass in each bin, presented in $\log(M/M_\odot\,h^{-1})$.
Best-fit values correspond to the maximum-a-posteriori (MAP) estimates, while the quoted uncertainties represent the 68\% projected joint-distribution highest posterior density (PJ-HPD) intervals.
The Probability To Exceed (PTE) is computed from the $\chi^{2}$ statistic, while the number of degrees of freedom is $\nu = 35$ for all bins. For each parameter, the first row shows the measured value with $1\sigma$ uncertainties, while the second row lists the corresponding input values from the MICE2 simulation.}
\end{table*}

\section{KiDS-DR5 Measurements}
\label{sec:kids}

\begin{table}
\caption{Stellar mass bins in KiDS-DR5 lens galaxy sample with summary statistics.}
\label{tab:kids_bins}
\centering
\begin{tabular}{lcccc}
\hline
Bin & Mass Range & $M_*^{\rm med}$&$z_{\rm B} ^{\rm med}$& $N_{\rm tot}$ \\
\hline
Bin 1 & (9.0,10.0] & 9.47 & 0.29 & 2278284 \\
Bin 2 & (10.0,10.4]& 10.18 & 0.32 & 545697 \\
Bin 3 & (10.4,10.7]& 10.53 & 0.35 & 257189 \\
Bin 4 & (10.7,11.0]& 10.82 & 0.39 & 142501 \\
Bin 5 & $>11$& 11.07 & 0.41 & 30883 \\
\hline
\end{tabular}
\tablefoot{$N_{\rm tot}$ gives the total number of galaxies in each bin. Stellar mass values are given in units of $\log(M_*/M_\odot\,h^{-1})$, while the superscript `med' denotes the median value of the sample.}
\end{table}

\begin{figure}
    \centering
    \includegraphics[width=0.9\linewidth]{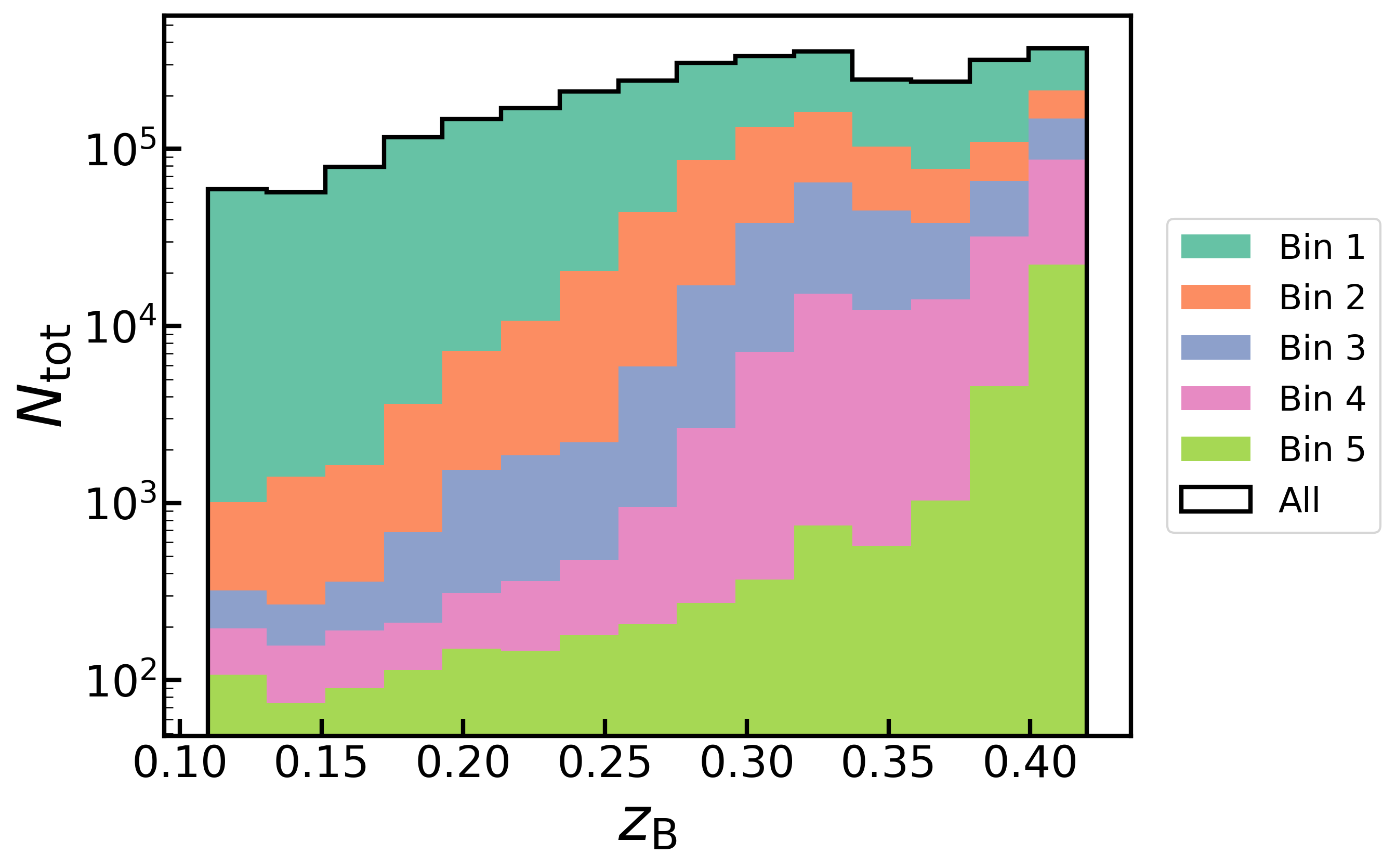}
    \caption{Photo-$z$ distributions of the lens galaxy samples in KiDS-DR5 in each stellar mass bin. The bins are plotted in order from highest to lowest redshift for visual clarity, but the bars are not cumulative totals; their heights directly reflect the number of galaxies in each bin. The log‑y axis is used to better show differences across bins with widely varying counts.}
    \label{fig:kids_nz_lens}
\end{figure}

\begin{figure}
    \centering
    \includegraphics[width=0.9\linewidth]{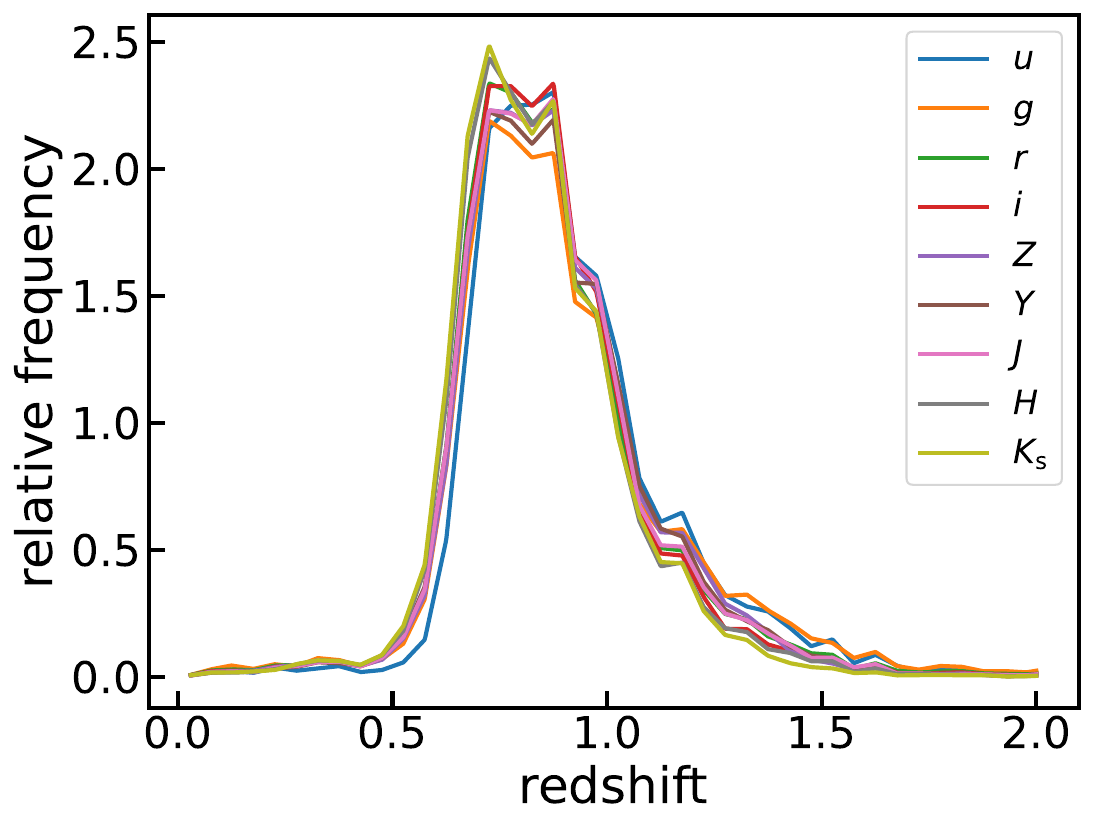}
    \caption{Redshift distributions of the source galaxy samples. Normalised redshift distributions $N(z)$ for KiDS-DR5 source galaxies selected with different magnitude cuts in the $ugriZYJHK_{\rm s}$ bands, estimated using SOMs. We find that the resulting distributions are in good agreement across bands and magnitude selections.}
    \label{fig:kids_nz}
\end{figure}

\begin{figure}
    \centering
    \includegraphics[width=0.9\linewidth]{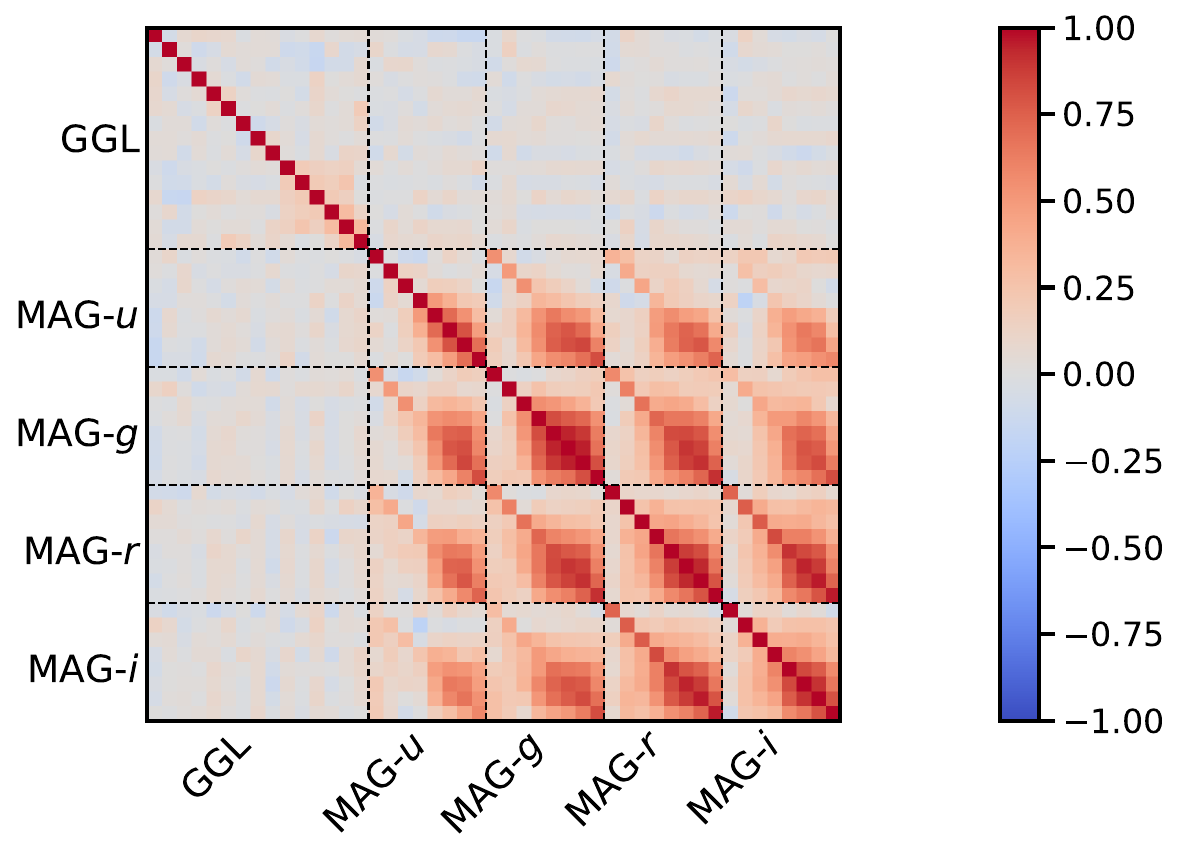}
    \caption{Representative correlation matrix of measurements with KiDS-DR5 data for the third stellar mass bin. As the matrices of other bins are similar, we present only this bin.
    }
    \label{fig:covmatrix}
\end{figure}

After validating our measurement and parameter inference framework with MICE2 simulations, we apply the same methodology to the final data release of the KiDS: KiDS-DR5. We select the lens and source galaxy samples following the same criteria adopted during our MICE2 tests: lens galaxies with $z_{\rm B}<0.42$, $m_r<23$ and stellar mass cuts defining five bins, and source galaxies with $z_{\rm B}>0.7$ and magnitude cuts in each band. The redshift gap between $0.4 < z_{\rm B} < 0.7$ is introduced to minimise spurious magnitude correlations arising from galaxies in the lens–source overlap, which do not originate from magnification or dust and would otherwise contaminate the signal. This gap ensures minimal contamination while retaining a high source number density. The stellar mass bin definitions and corresponding summary statistics of the KiDS-DR5 lens sample are listed in Table~\ref{tab:kids_bins}. The photometric redshift distributions of the lenses in each stellar mass bin are shown in Fig.~\ref{fig:kids_nz_lens}, illustrating the redshift coverage and relative number densities across the bins.

In KiDS-DR5, the source population extends to higher redshifts, reaching $z_{\rm B} \sim 2.0$, allowing us to use background galaxies beyond the maximum redshift of the MICE2 simulation. Additionally, we restrict both the lens and source samples to objects classified as galaxies (SG2DPHOT$=0$) to avoid contamination from stars.
The source redshift distributions $N(z)$ for each band are shown in Fig.~\ref{fig:kids_nz}. These distributions are computed using self-organising maps (SOMs) to estimate the true redshift distribution from the photometric redshifts. Despite the different magnitude cuts applied for magnitude shift measurements in each band, the resulting $N(z)$ distributions are highly similar across bands. Consequently, the critical surface density $\Sigma_{\rm crit}$, differs by less than 0.5\% between bands, justifying the use of the $N(z)$ with $r$-band cut for $\Sigma_{\rm crit}$ in the rest of the analysis. While the source selection is anchored in the $r$-band, the application of band-dependent magnitude cuts leads to non-identical samples, such that the observed consistency of the resulting $N(z)$ distributions across bands is not a priori guaranteed.

In addition to the optical measurements, we also perform the magnitude shift analysis in the NIR bands $Z Y J H K_{\rm s}$. We find that the SOM-estimated redshift distributions of the source samples selected in the NIR bands are consistent with those of the optical samples. Since no significant differences are observed, we adopt the same $r$-band $N(z)$ for the computation of $\Sigma_{\rm crit}$ for the NIR measurements as well.

We performed all measurements with \textsc{TreeCorr}, following the same procedure adopted in the MICE2 tests. Random subtraction is applied to remove systematic contributions not associated with the lens-source correlation. Magnitude shift signals are measured in eight logarithmically spaced angular bins between 0.1 and 60 arcmin, while GGL signals are measured in 15 bins over the same range. Unlike in MICE2, KiDS-DR5 allows reliable measurements below 1 arcmin, enabling the use of additional GGL bins on small scales.

Having measured the signals of magnitude shift and GGL in KiDS-DR5 data, we proceed to their physical interpretation. As in the MICE2 tests, we use a model that simultaneously describes the matter distribution in halo and the dust distribution around lens galaxies, allowing us to interpret the observed GGL and magnitude shift signals in a consistent framework. 

The modelling of the KiDS-DR5 measurements follows the same framework adopted in the MICE2 analysis. The extinction profile is computed in the rest frame of the lenses and converted to the observer frame using a wavelength-dependent extinction law. We adopt the \citet{fitzp_2019} extinction curve with $R_V = 3.1$ and evaluate the extinction amplitude at the effective wavelengths of the KiDS optical bands ($u g r i$). We note that the KiDS-DR5 dataset provides optical magnitudes in five bands, $u g r i_1$ and $i_2$. We measure the magnitude shift signal independently in both the $i_1$ and $i_2$ bands and find them to be statistically indistinguishable, as expected given their identical effective wavelengths. For the remainder of the analysis, we therefore adopt as our $i$-band measurement the average of the $i_1$ and $i_2$ signals. This approach preserves the available information while avoiding unnecessary duplication in the covariance estimation.

\begin{table*}
\centering
\caption{Best-fit parameters with $1\sigma$ uncertainties measured through joint analysis of magnitude shift and GGL in KiDS-DR5.}
\begin{tabular}{lccccc}
\toprule
Parameter & Bin 1 & Bin 2 & Bin 3 & Bin 4 & Bin 5 \\
\midrule
$\log(M_{\mathrm{c}})$ & $11.66^{+0.12}_{-0.20}$ & $11.95^{+0.09}_{-0.14}$ & $12.17^{+0.12}_{-0.11}$ & $12.38^{+0.09}_{-0.13}$ & $12.63^{+0.12}_{-0.09}$ \\ [0.4em]
$\log(M_{\mathrm{d1}})$ & $7.53^{+0.16}_{-0.27}$ & $7.52^{+0.09}_{-0.10}$ & $7.69^{+0.11}_{-0.10}$ & $7.77^{+0.08}_{-0.13}$ & $8.02^{+0.09}_{-0.13}$ \\ [0.4em]
\midrule
$\chi^{2}\;(\nu=40)$ & 51.19 & 47.48 & 47.79 & 41.90 & 43.62 \\
PTE & 0.11 & 0.19 & 0.18 & 0.39 & 0.32 \\
\bottomrule
\end{tabular}
    \tablefoot{Parameters are given in $\log(M/M_\odot\,h^{-1})$. Best-fit values correspond to the MAP estimates, while the quoted uncertainties represent the 68\% PJ-HPD intervals. The number of degrees of freedom is $\nu = 40$ for all stellar mass bins.}
\label{tab:kids_bestfit}
\end{table*}

Parameter inference is performed using the same MCMC setup as in the MICE2 tests, employing the \textsc{emcee} ensemble sampler to jointly fit the multi-band magnitude-shift and GGL measurements. The model includes the same seven free parameters and uniform priors listed in Table~\ref{tab:priors}, and we assume a Gaussian likelihood with the full jackknife covariance matrix of the measurements. Covariance matrices are estimated using the \textsc{TreeCorr} jackknife implementation. Given the large number of jackknife regions (1300) and the relatively small scales probed, the jackknife estimate is expected to provide a robust approximation of the covariance. As an illustration, Fig.~\ref{fig:covmatrix} shows the correlation matrix for the third stellar-mass bin; the covariance matrices of the remaining bins are qualitatively similar.

In Table~\ref{tab:kids_bestfit}, we present the best-fit results from the KiDS–DR5 analysis, analogous to those shown for the MICE2 tests in Table~\ref{tab:results_mice2}. Best-fit values correspond to the MAP estimates, and quoted uncertainties again represent the 68\% PJ–HPD intervals. The reported PTE values indicate that the joint model provides a good description of the observed signals in all stellar-mass bins.

\begin{figure*}
    \centering
    \includegraphics[width=0.9\textwidth]{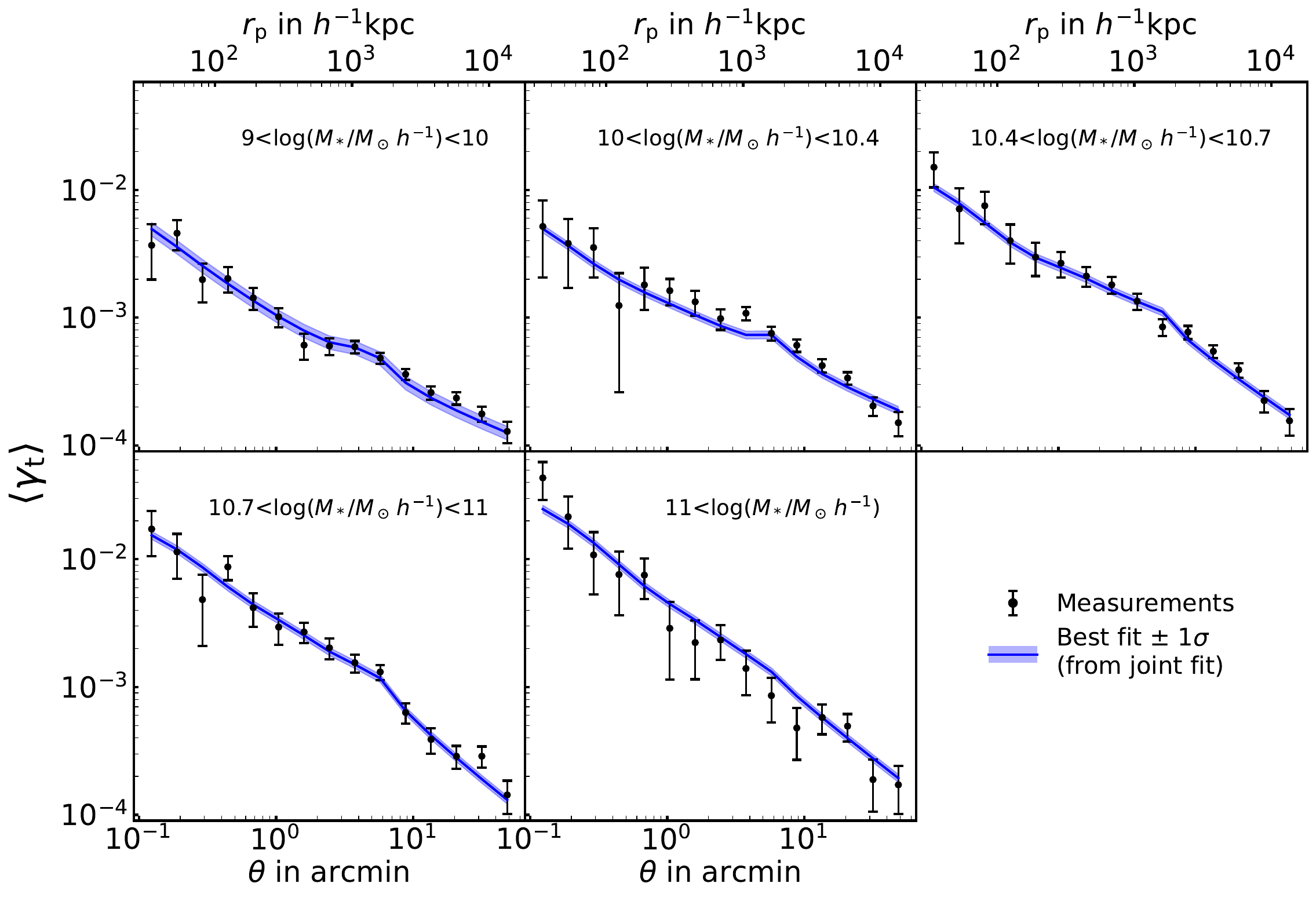}
    \caption{GGL signals in stellar mass bins measured in KiDS-DR5, alongside the best fit with $1\sigma$ posterior credible interval from the joint fit with the magnitude shift. Physical separations are computed using the redshift of $z\sim0.35$.
    }
    \label{fig:kids_ggl}
\end{figure*}

Figure~\ref{fig:kids_ggl} presents the GGL measurements for the five stellar mass bins, each shown together with the best-fitting model and its $1\sigma$ confidence interval derived from the joint analysis of GGL and magnitude shift signals. The overall agreement between the data and the model indicates that the inferred halo masses and satellite parameters provide a consistent description of the observed lensing signal across all mass bins.

\begin{figure*}
    \centering
    \includegraphics[width=0.9\textwidth]{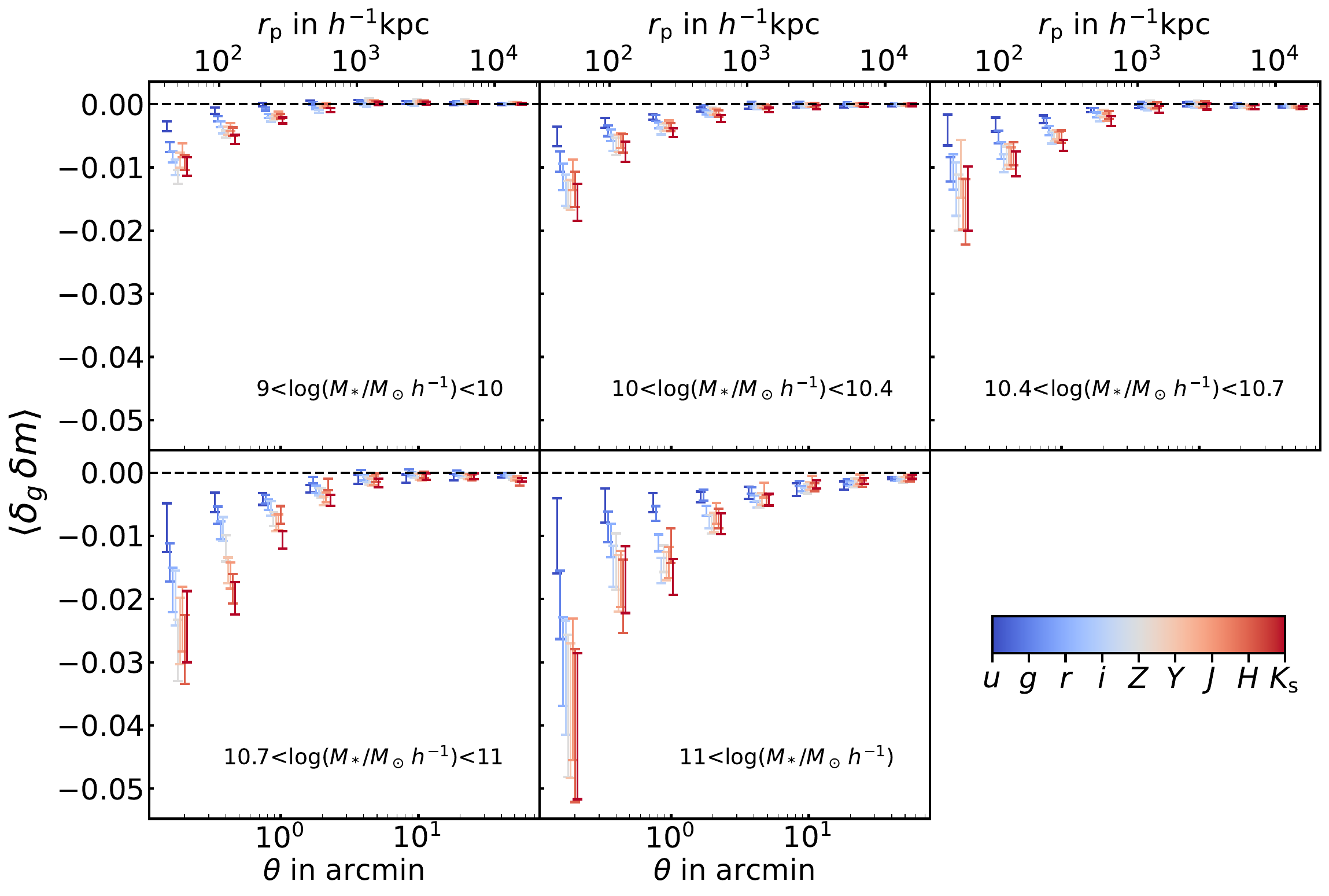}
    \caption{Magnitude-shift signals in stellar mass bins measured in KiDS-DR5 across $ugriZYJHK_\mathrm{s}$ bands, showing measurements only without model fits.  Physical separations are computed using the redshift of $z\sim0.35$.
    }
    \label{fig:kids_magn}
\end{figure*}

The corresponding magnitude shift measurements are shown in Fig.~\ref{fig:kids_magn}. A clear stellar-mass dependence is visible in the longer-wavelength bands, where magnification is the dominant contribution. In contrast, at shorter wavelengths — most notably in the $u$ and $g$ bands — the ordering with stellar mass becomes less pronounced. This reflects the increasing impact of dust extinction at short wavelengths, where the total magnitude shift is no longer governed primarily by magnification but receives a substantial chromatic contribution from dust.

As a consequence, higher stellar (or halo) mass does not necessarily correspond to a larger magnitude shift amplitude in the $u$- and $g-$ bands, since variations in dust content can compensate for, or even dominate over, the lensing magnification signal. In contrast, in the near-infrared bands, where dust extinction effects are negligible, the magnitude shift signal is dominated by magnification and exhibits a clear hierarchical increase with stellar mass. This wavelength-dependent behaviour highlights the importance of multi-band measurements for disentangling dust extinction from magnification and for robustly inferring both halo and dust properties.

\begin{figure*}
    \centering
    \includegraphics[width=0.9\textwidth]{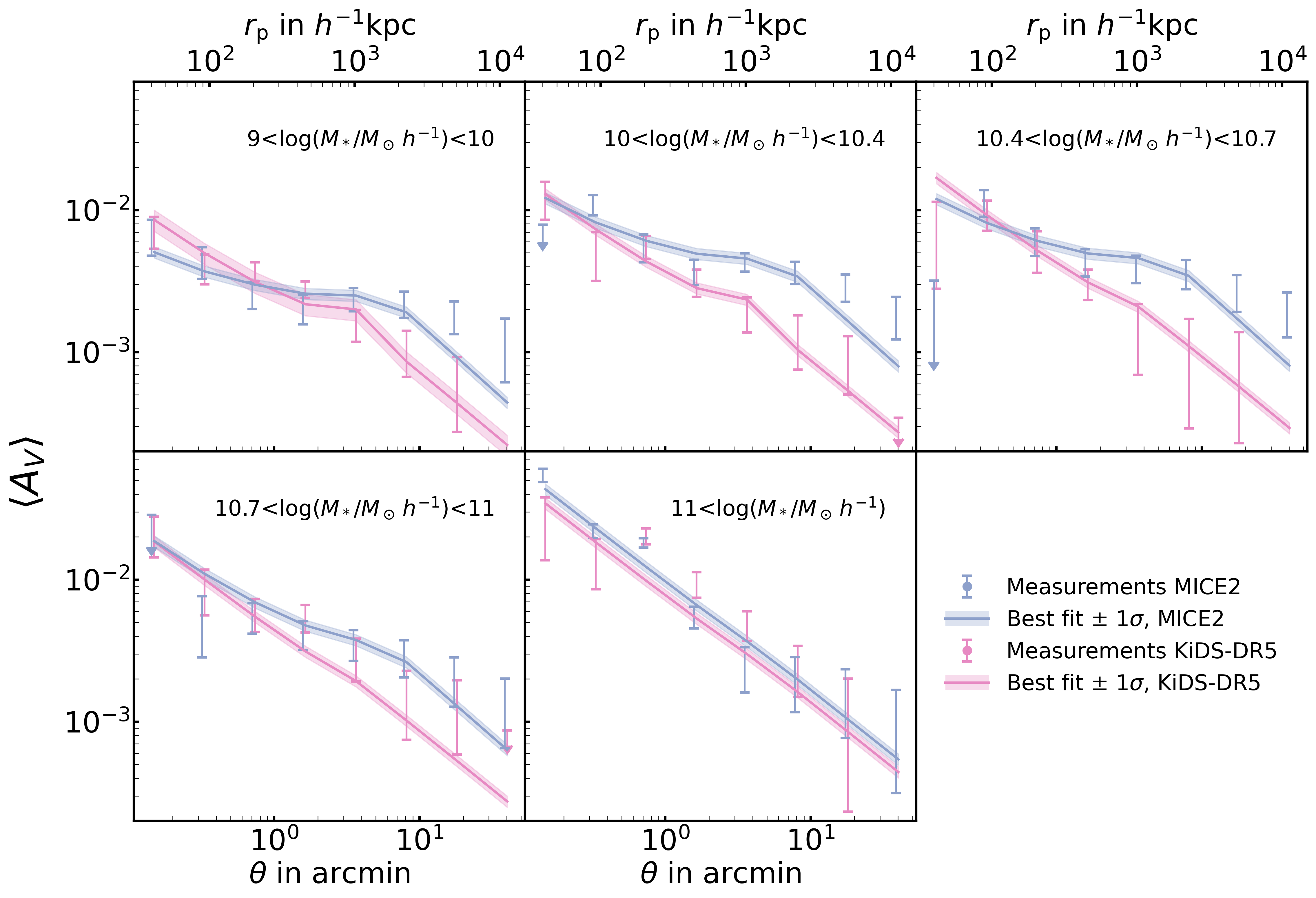}
    \caption{Rest-frame extinction profiles as a function of projected distance for different stellar-mass bins. 
Purple data points show measurements from the MICE2 simulation, while pink data points correspond to KiDS–DR5 observations. 
The profiles are computed as covariance-weighted averages over optical ($u g r i$) and near-infrared ($Z Y J H K_s$) band combinations, converted to extinction.  Physical separations are computed using the redshift of $z\sim0.35$.}
    \label{fig:mice_vs_kids_Av}
\end{figure*}

\begin{figure*}
    \centering
    \includegraphics[width=0.9\textwidth]{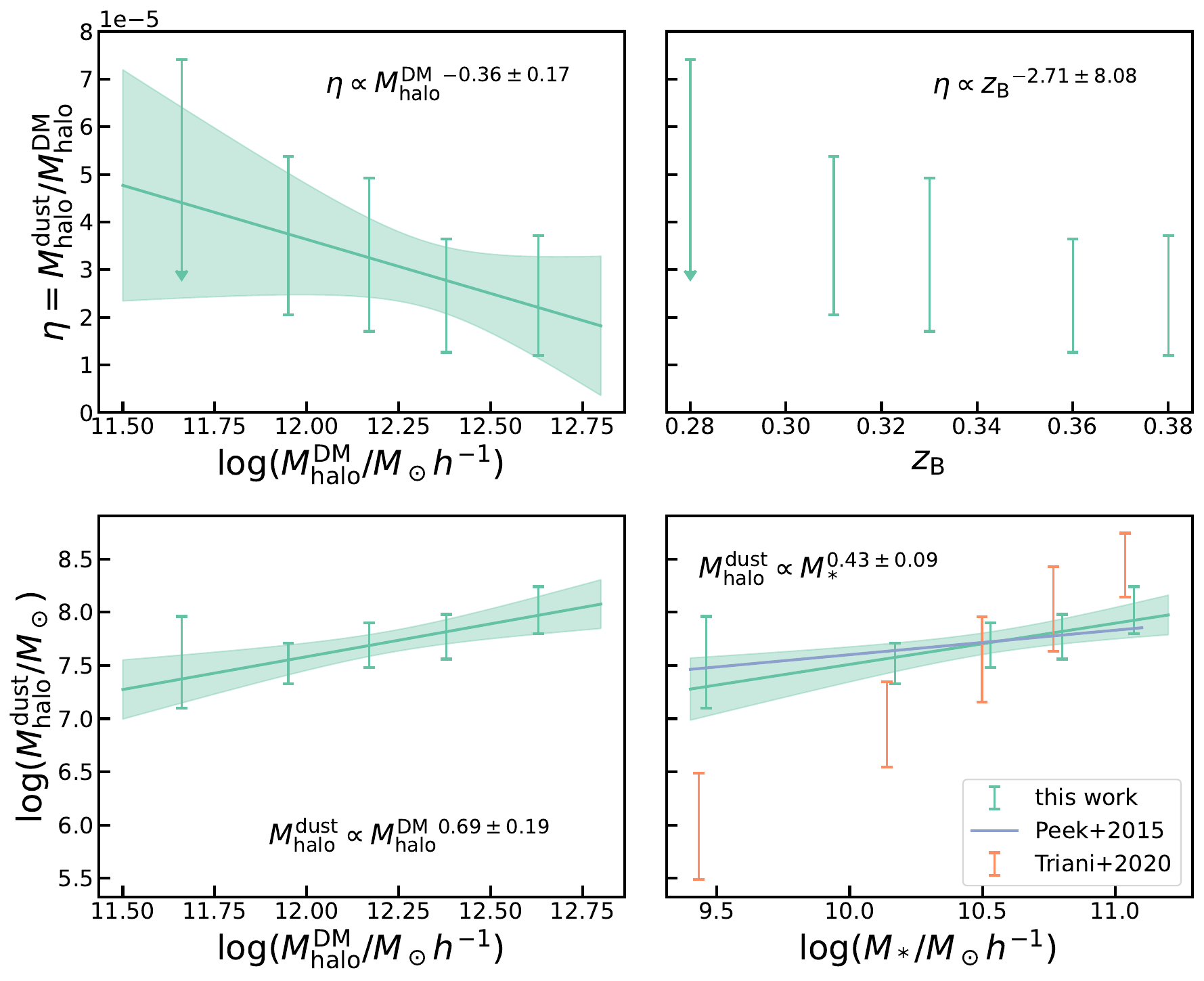}
    \caption{Summary of inferred dust properties of analysed KiDS–DR5 lens galaxy sample. 
\textit{Upper left:} $\eta$ as a function of halo mass. No statistically significant trend is apparent due to the relatively large uncertainties. \textit{Upper right:} $\eta$ as a function of redshift, no clear redshift evolution is detected within the current errors. \textit{Lower left:} Dust mass versus halo mass, showing a clear power-law scaling $M_{\rm dust} \propto M_{\rm halo}^{0.69}$, indicating that more massive halos host larger amounts of circumgalactic dust. 
\textit{Lower right:} Dust mass as a function of stellar mass, revealing the expected correlation with galaxy stellar content. We also compare our results with the semi-analytic model results of \citet{triani_2020} and observational results of \citet{peek_2015}.
Error bars indicate $1\sigma$ uncertainties in each measurement.}

    \label{fig:dtm_z}
\end{figure*}

In Fig.~\ref{fig:mice_vs_kids_Av} we present the rest-frame extinction profiles for each stellar-mass bin, measured both in the MICE2 simulation and the KiDS–DR5 data. The profiles are computed using Eq.\eqref{eq:weighted_A_V}. Fig.~\ref{fig:dtm_z} summarises the inferred dust properties of KiDS–DR5 lens galaxies. Error bars represent propagated $1\sigma$ uncertainties from the MCMC analysis. The shaded regions indicate the $1\sigma$ uncertainty of the best-fit scaling relations. The upper-left panel shows the dust-to-mass ratio $\eta$ as a function of halo mass, while the upper-right panel shows $\eta$ as a function of redshift. We find evidence for a scaling of $\eta$ with halo mass, $\eta \propto M_{\rm halo}^{-0.36 \pm 0.17}$, indicating a mild decrease of the dust-to-mass ratio towards higher halo masses. In contrast, the dependence of $\eta$ on redshift is not statistically significant, with $\eta \propto z_{\rm B}^{-2.71 \pm 8.08}$, where the large uncertainty on the slope prevents a robust detection of any trend. Within the current uncertainties, the redshift evolution of $\eta$ remains unconstrained.
The lower panels show the absolute dust mass scaling relations. The lower left panel presents $M_{\rm dust}$ versus $M_{\rm halo}$, while the lower right panel shows $M_{\rm dust}$ versus $M_*$. Data points follow a power-law scaling $M_{\rm dust} \propto M_*^{0.43\pm0.09}$. For $M_{\rm dust}$ versus $M_{\rm halo}$, the best-fit power law is $M_{\rm dust} \propto M_{\rm halo}^{0.69\pm0.19}$. 

For comparison, the lower right panel also includes literature relations. The dust-to-stellar mass relation from the observational study of \citet{peek_2015}, based on low-redshift ($z<0.1$) galaxies, is shown as a solid line. In addition, we include results from the semi-analytic galaxy evolution model \textsc{Dusty-SAGE} \citep{triani_2020}, for which we apply the same stellar-mass binning and selection criteria (including magnitude and redshift cuts) as used in our KiDS–DR5 analysis, and plot the mean dust mass in each bin. Overall, our measurements are in good agreement with \citet{peek_2015} and with \textsc{Dusty-SAGE} \citep{triani_2020} for massive systems, although some differences in the slope of the dust–to-stellar mass relation are found, which we discuss in Sect.~\ref{sect:disc}.

\section{Discussions}
\label{sect:disc}
While the overall amplitudes of the extinction signals are broadly consistent between the KiDS–DR5 and MICE2, the MICE2 profiles are noticeably flatter on large scales in the three lowest stellar mass bins. This behaviour can be directly linked to the inferred satellite fractions. In the MICE2 simulations, the lowest stellar mass bins are dominated by satellite galaxies, with measured satellite fractions of $f_{\rm sat}\sim0.8-0.9$. In this regime, the extinction profile measured on large scales (e.g. $>300\,h^{-1}$ kpc) is largely driven by dust associated with neighbouring host halos, leading to a shallower radial dependence. In contrast, the KiDS-DR5 measurements yield systematically lower satellite fractions in the same stellar mass bins ($f_{\rm sat}\sim0.6-0.7$), resulting in extinction profiles that are more centrally concentrated and less affected by host contributions at large radii.

The scaling relations of dust mass with halo and stellar mass reveal clear trends. More massive halos host systematically larger amounts of circumgalactic dust. The observed scaling of dust mass with both halo mass and stellar mass suggests that the circumgalactic dust content is closely linked to the overall depth of the gravitational potential and the integrated star formation history of galaxies. The sub-linear slopes of the scaling relations are consistent with a scenario in which dust production in galaxies is counterbalanced by dust destruction and removal processes in massive halos, such as thermal sputtering and grain–grain collisions in hot circumgalactic gas.

Our inferred dust–stellar mass relation is consistent, within the $1\sigma$ uncertainties, with the results of \citet{peek_2015}, who found $M_\mathrm{dust} \propto M_*^{0.23}$. While our best-fit slope appears moderately steeper, the result from \citet{peek_2015} lies within the uncertainty range of our measurement, as shown by the shaded confidence region in Fig.~\ref{fig:dtm_z}. We attribute the difference in the best-fit values to the different redshift regimes probed. While \citet{peek_2015} focused on a very local galaxy population ($z<0.1$), our analysis spans a wider redshift range ($0.1 \lesssim z \lesssim 0.4$). Considering that the efficiency of dust production, retention, and transportation evolves with cosmic time, for example due to changes in star formation activity, feedback strength, or dust survival in galactic halos, averaging over a broader redshift range can lead to an effective dust–stellar mass relation that differs from measurements confined to a narrow, local redshift interval. In this sense, the steeper relation inferred here may partly reflect the inclusion of galaxies spanning a wider range of evolutionary stages.

In contrast, the predictions from the \textsc{Dusty-SAGE} semi-analytic model exhibit a substantially steeper dust–stellar mass relation. This behaviour indicates that, within the \textsc{Dusty-SAGE} framework, low-mass galaxies are predicted to host significantly less circumgalactic dust compared to more massive systems. Such a steep scaling is likely driven by the specific feedback prescriptions implemented in the model. 
In particular, efficient supernova feedback in low-mass halos can expel metals and dust from the interstellar medium, while strong AGN feedback in massive systems regulates dust growth and retention at the high-mass end. In \textsc{Dusty-SAGE}, dust is efficiently destroyed in galactic winds driven by supernovae, whereas other models \citep[e.g., ][]{richie_2024} suggest that a fraction of this dust can survive.
Additionally, the treatment of dust destruction by shocks and thermal sputtering, as well as the assumed efficiencies of dust growth and transport into the circumgalactic medium, can further suppress the dust content of low-mass galaxies. These modeling choices could collectively steepen the predicted dust–to-stellar mass relation relative to our observationally inferred result.

\section{Conclusions}
\label{sec:conclusions}
In this work, we have measured the circumgalactic dust content of a well-defined sample of galaxies in the KiDS–DR5 data, building on a methodology that was validated using MICE2 simulations in our previous study. Our primary aim was to characterise how CGM dust properties scale with stellar mass, and to establish observational relations that can serve as empirical benchmarks for models of dust transport and survival in galaxy halos.

We first validated our measurement pipeline with MICE2 simulations, demonstrating that our combined multi-band magnitude shift and GGL framework reliably recover both mean halo masses and dust masses across a range of stellar mass bins. During this validation, we found that accurately modelling the distinct contributions of central and satellite galaxies is essential, particularly in low-stellar-mass bins where the satellite fraction is high. This motivated an extension of our model to explicitly include the satellite fraction and the off-centred host-halo contribution following \citet{sifon_2015}. The MICE2 tests further showed that our modelling framework captures the statistical properties of the measured signals well. As a result, the input halo and dust masses are recovered within $1\sigma$ in all stellar mass bins except the lowest-mass bin, where the dust mass is recovered within $2\sigma$.

Applying the same methodology to KiDS-DR5, we measured magnitude shifts in the $ugri_1$ bands and GGL signals in five stellar mass bins. Due to relatively large uncertainties, the dust-to-mass ratio as a function of halo mass and redshift could not be robustly constrained. However, we detect clear scaling relations between dust mass and both stellar and halo mass, approximately following $M_{\rm dust} \propto M_{\rm halo}^{0.69}$ and $M_{\rm dust} \propto M_*^{0.43}$ for all bins above the lowest stellar-mass range. The lowest-mass bin shows deviations from this trend, likely driven by its broader stellar-mass selection. Although the median redshift varies across the stellar-mass bins, the substantial redshift overlap and intrinsic scatter within each bin ($\sigma_z > 0.07$) make it unlikely that the observed dust–stellar mass relation is primarily driven by redshift evolution. We therefore interpret the measured scaling as reflecting a genuine dependence of circumgalactic dust mass on stellar mass over the redshift range probed by the KiDS–DR5 sample.

Our analysis highlights several limitations. While the KiDS–DR5 data provide sufficient statistical power to detect scaling relations, deeper imaging or a higher density of background sources would significantly improve the S/N, particularly for low-mass galaxies, and enable tighter constraints on the dust-to-mass ratio and its redshift dependence. Moreover, degeneracies between observed galaxy properties, e.g. stellar mass and redshift, cannot be fully disentangled with the current data. As a result, the measured scaling relations should be interpreted as effective trends averaged over the joint distribution of these properties.

We further assume a fixed SMC-type dust and a fixed slope of $-0.8$ for the radial dust extinction profile, following \citet{menard_2010}, corresponding to a constant extinction coefficient $K_{\rm ext}$ and a uniform dust grain composition. In reality, dust properties may vary with galaxy mass, environment, or redshift, introducing additional complexity not captured by our model. In addition, weak-lensing–based measurements inherently probe only the mean properties of a galaxy sample, rendering the analysis insensitive to outliers or intrinsic scatter within individual bins. Finally, our lens redshift range is limited by photometric redshift uncertainties; to minimise source–lens overlap, we restrict the analysis to $z_{\rm B} < 0.4$, preventing an extension to higher redshift galaxies.

Taken together with previous observational and theoretical studies, such as the low-redshift measurements of \citet{peek_2015} and the steeper relations predicted by the \textsc{Dusty-SAGE} model \citep{triani_2020}, our results underscore the importance of probing circumgalactic dust over a wide range of stellar masses, luminosities, and redshifts in order to place meaningful constraints on models of dust production, transport, and survival in galaxy halos.

Our measurements provide new observational benchmarks for semi-analytic and hydrodynamical simulations, offering guidance for feedback modelling in galaxy evolution studies. Future surveys with higher source densities and deeper imaging, such as \textit{Euclid} \citep{euclid_2025}, will allow the extension of this methodology to larger volumes, finer stellar mass bins, and higher redshifts, ultimately providing more stringent constraints on the lifecycle of dust in and around galaxies.

\begin{acknowledgements}
       We would like to thank Dillon Brout and Brice Ménard for useful discussions during the project. EG acknowledges the support from the Deutsche Forschungsgemeinschaft (DFG) SFB1491. AHW is supported by the Deutsches Zentrum für Luft- und Raumfahrt (DLR), made possible by the Bundesministerium für Wirtschaft und Klimaschutz, under project 50QE2305, and acknowledge funding from the German Science Foundation DFG, via the Collaborative Research Center SFB1491 "Cosmic Interacting Matters - From Source to Signal". H. Hildebrandt is supported by a DFG Heisenberg grant (Hi 1495/5-1), the DFG Collaborative Research Center SFB1491, an ERC Consolidator Grant (No. 770935), and the DLR project 50QE2305.
       Marika Asgari acknowledges the UK Science and Technology Facilities Council (STFC) under grant number ST/Y002652/1 and the Royal Society under grant numbers RGSR222226 and ICAR1231094. BJ acknowledges support by the ERC-selected UKRI Frontier Research Grant EP/Y03015X/1 and by STFC Consolidated Grant ST/V000780/1.  SJ acknowledges the Ramón y Cajal Fellowship (RYC2022-036431-I) from the Spanish Ministry of Science.
       HHo acknowledges support from the European Research Council (ERC) under the European Union's Horizon 2020 research and innovation program with Grant agreement No. 101053992. LM acknowledges the financial contribution from the grant ASI n. 2024-10-HH.0 ``Attività scientifiche per la missione Euclid – fase E''. RR is partially supported by an ERC Consolidator Grant (No. 770935). The data used in this work are based on observations made with ESO Telescopes at the La Silla Paranal Observatory under programme IDs 177.A-3016, 177.A-3017, 177.A-3018 and 179.A-2004, and on data products produced by the KiDS Consortium.  The KiDS production team acknowledges support from the DFG, ERC, NOVA and NWO-M grants; Target; the University of Padova, and the University Federico II (Naples). This work was supported by a grant of the German Centre of Cosmological Lensing, hosted at Bochum University. 

\end{acknowledgements}

%
%

\bibliographystyle{aa}
\bibliography{References}

\appendix
\section{Sensitivity of the extinction profile to model parameters}
\label{app:Av_fsat_Rsat}

\begin{figure*}
    \centering
    \includegraphics[width=0.9\textwidth]{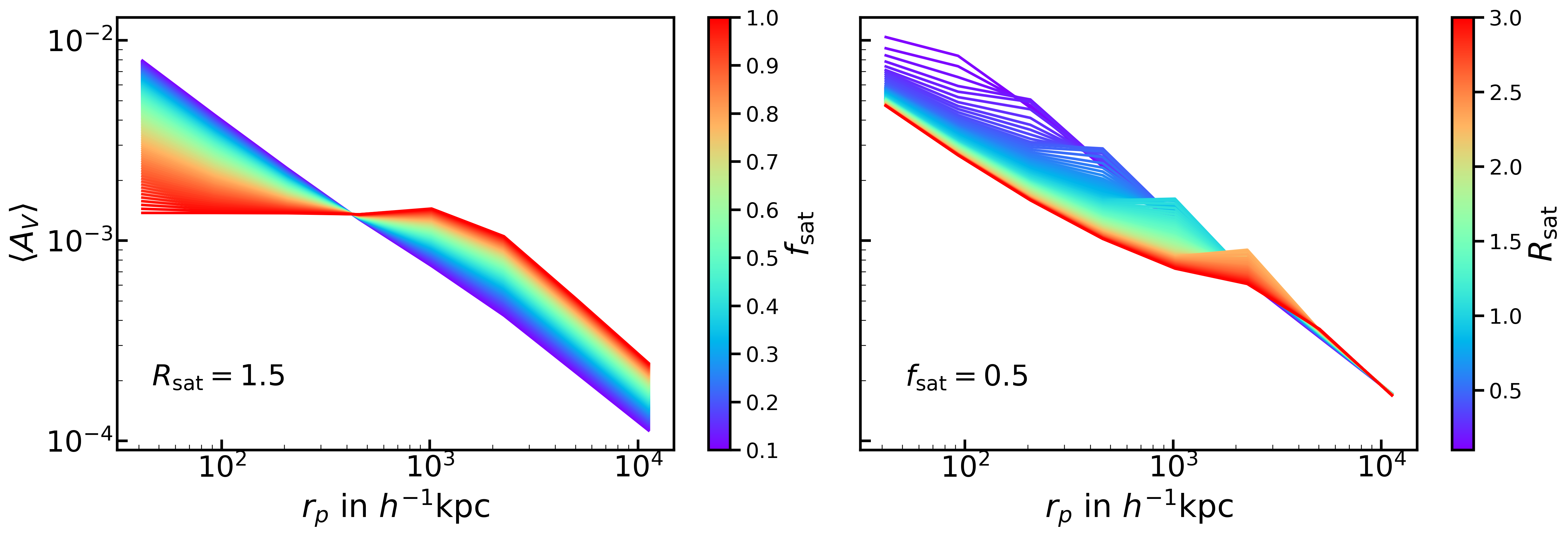}
    \caption{Extinction profiles in the $V$ band as a function of projected separation for varying model parameters. Left: Profiles for different satellite fractions $f_\mathrm{sat}$, with other parameters fixed. Higher satellite fractions flatten the profile, enhancing the signal on large scales. Right: Profiles for different satellite scale radii $R_\mathrm{sat}$, with other parameters fixed. The peak of the profile shifts to scales corresponding to $R_\mathrm{sat}$, reflecting the average host–satellite separation. Colour gradients indicate the parameter values.}
    \label{fig:Av_fsat_Rsat}
\end{figure*}
In Fig.~\ref{fig:Av_fsat_Rsat}, we illustrate the sensitivity of our extinction model in the $V$ band to the satellite fraction ($f_\mathrm{sat}$) and the satellite scale radius ($R_\mathrm{sat}$).
The left panel shows the extinction profile as a function of projected separation for varying $f_\mathrm{sat}$ while keeping all other parameters fixed. Increasing the satellite fraction tends to flatten the profile: higher $f_\mathrm{sat}$ implies a stronger contribution from host galaxies offset from satellites, which boosts the signal on large scales while suppressing it in the small-scale (1-halo) regime.
The right panel shows the profile for varying $R_\mathrm{sat}$ with other parameters held constant. The profile peaks roughly at the value of $R_\mathrm{sat}$, as expected, since this parameter represents the average distance between host galaxy centers and satellites. Smaller $R_\mathrm{sat}$ leads to satellites being closer to their hosts, resulting in a stronger contribution to the profile at smaller scales.

\end{document}